\UseRawInputEncoding
\documentclass[aps,prb,twocolumn,superscriptaddress,longbibliography,nobalancelastpage]{revtex4-2}
    
    \usepackage{amsmath}
    \usepackage{graphicx}
    \usepackage{physics}
    \usepackage{comment}
    \usepackage{hyperref}
    \usepackage{soul,color}
    \usepackage{bbold}
    \usepackage{xcolor}

    \newcommand{\da}{\downarrow}
    \newcommand{\ua}{\uparrow}
    \newcommand{\ad}{\hat{a}^\dagger}
    \newcommand{\an}{\hat{a}^{}}
    \newcommand{\bd}{\hat{b}^\dagger}
    \newcommand{\bn}{\hat{b}^{}}
    \renewcommand{\sd}{\hat{\sigma}^{+}}
    \newcommand{\sn}{\hat{\sigma}^{-}}

    \newcommand{\Da}{\Downarrow}
    \newcommand{\Ua}{\Uparrow}
    
    \newcommand{\td}{\tilde\delta}
    
    \newcommand{\symset}{\mathcal{S}}
    \newcommand{\cntset}{\mathcal{P}} 
    \newcommand{\indset}{\mathcal{I}} 
    
    \newcommand{\mapto}{\mathcal{M}}
    \newcommand{\maptop}{\mapto_{p-1 \mapsto p}}
    \newcommand{\maptonp}{\mapto_{N-p \mapsto N-p+1}}
    \newcommand{\maptonpm}{\mapto_{N-p-1 \mapsto N-p}}

    \newcommand{\matelem}{M}
    
    \newcommand{\maxnu}{\nu_{\text{max}}}
    
    \newcommand{\absorb}{A}
    \newcommand{\hopresp}{M}
    
    \newcommand{\CountSigma}{N}
    \newcommand{\CountFTSigma}{\tilde{N}}

    \newcommand{\aset}{\mathcal{A}}

    \newcommand{\wv}{{\omega_{v}}}
    
    \newcommand{\lam}{\lambda}
    \newcommand{\lamp}{\lambda_D}
    \newcommand{\wh}{\omega_D}
    \newcommand{\wo}{\omega_0}
    \newcommand{\wrr}{\omega_R}
    
    \newcommand{\twr}{\wrr\sqrt{\rho_{ex}}}
    
    \newcommand{\kD}{\ket{D}}
    \newcommand{\kup}{\ket{\uparrow}}
    \newcommand{\kdn}{\ket{\downarrow}}

    \newcommand{\com}[1]{}

    \graphicspath{{/pdf/}}
    \usepackage{float}
    
\begin{document}
    
    \author{M. Ahsan Zeb}
    \affiliation{Department of Physics, Quaid-i-Azam University, Islamabad 45320, Pakistan}
    \affiliation{SUPA, School of Physics and Astronomy, University of St Andrews, St Andrews, KY16 9SS, United Kingdom}
    \author{Peter G. Kirton}
    \affiliation{Department of Physics and SUPA, University of Strathclyde, Glasgow G4 0NG, United Kingdom}
    \affiliation{Vienna Center for Quantum Science and Technology, Atominstitut, TU Wien, 1040 Vienna, Austria}
    \affiliation{SUPA, School of Physics and Astronomy, University of St Andrews, St Andrews, KY16 9SS, United Kingdom}
    \author{Jonathan Keeling}
    \affiliation{SUPA, School of Physics and Astronomy, University of St Andrews, St Andrews, KY16 9SS, United Kingdom}
    \date{\today}
    
    \title{Incoherent charge transport in an organic polariton condensate}
    \begin{abstract}
    We study how polariton condensation modifies charge transport in organic materials.  In typical organic materials, charge transport proceeds via incoherent hopping. We therefore provide an approach to determine how the rate and final state of this hopping process is affected by strong matter-light coupling and polariton condensation.  
    We show how the hopping process may create excitations when starting from a state with a finite excitation density.  That is, how hopping can change the state of a lower polariton condensate by creating
    upper polaritons, optically inactive excitonic dark states, or by exciting vibrational sidebands. 
    While the matrix elements for these processes can be large, for typical materials at room temperature, such excitations are suppressed by thermal factors, and ground state processes dominate.   
    We thus study how the ground state hopping rate depends on condensate density, matter-light coupling, and cavity photon detuning. 
    All these factors change the vibrational configuration associated with the optically active molecules, which can enhance or suppress hopping by increasing or decreasing the vibrational overlap with the state of a charged molecule.
    We show that hopping rates can be exponentially sensitive to detuning and condensate density, allowing an increase or decrease of hopping rate by two orders of magnitude.
    \end{abstract}
    
    \maketitle
    
    \section{Introduction}
    
    In organic light emitting devices, charge transport is an incoherent process of hopping between molecules~\cite{SchmidlinPMB80,Bassler99a,Bassler99b,Pope}.
    Understanding how such transport is affected by material properties---such as disorder and vibrational dressing---is crucial to enable design of more efficient light emitting and light harvesting materials.
    Many organic materials also show large oscillator strengths, so can reach the strong matter-light coupling regime~\cite{lidzey98,lidzey99,Lidzey00,holmes2004,Tischler2005}.
    Strong coupling changes the energies and nature of molecular eigenstates, and can thus influence transport.
    In this paper, we discuss how strong matter-light coupling affects incoherent hopping transport in the presence of a polariton condensate.

    The effects of strong matter-light coupling on material properties have been extensively studied. This includes experiments~\cite{Hutchison12,thomas2016,munkhbat2018suppression,Thomas2019} and theory~\cite{HerreraPRL16,Galego2016,galego17,MartinezMartinez2018a,Du21,li2021cavity,schafer2021shining,Yang2021,li2021collective,Chowdhury21, mandal2021theory,Pannir22,du2021dark} considering how chemical reaction rates can be changed (reviewed in Refs.~\cite{ebbesen2016hybrid,feist2017polaritonic,Ribeiro2018,GarciaVidal2021Review,Wang21,Nagarajan21}), and work on changing the superconducting transition temperature~\cite{Sentef2018,thomas2019exploring}, building on experiments on light-induced superconductivity~\cite{Fausti2011,Mankowsky2014,Mitrano2016,Schlawin2017}.
    The effects of strong matter-light coupling on transport have also been explored both
    experimentally~\cite{Orgiu2015} and theoretically~\cite{Feist2015,Schachenmayer2015,HerreraPRL16,Hagenmueller2017,Hagenmuller2018,Schafer2019,botzung2020:dark,Wellnitz21}, including ballistic and incoherent charge transport, as well as energy transport.  
  
    The focus of this paper is  on the combined effect of strong matter-light coupling and polariton condensation on hopping transport.  
    Polariton condensation~\cite{Kasprzak2006,Balili2007,Carusotto2013} refers to a state with a single macroscopically occupied polariton mode.  In thermal equilibrium this is akin to Bose--Einstein condensation. With finite polariton lifetime it is closer to a laser, but with stimulated emission replaced by stimulated scattering. Polariton condensation has been seen in many organic materials~\cite{Forrest2010,Daskalakis2014,Plumhof14,grant2016,cookson2017,dietrich16,betzold2019,Rajendran2019,Wei2019}; for a review, see~\cite{Keeling2020:review}.
    In most cases, polariton condensation is driven by optical (i.e.~external laser) pumping, while electrical pumping has been realized with inorganic materials~\cite{schneider2013,Bhattacharya2013}.
    Some questions about the interaction between a polariton condensate and charge transport have been considered for inorganic polariton condensates~\cite{Myers2018,Cotlet2018,Chervy2020,Li2021}, where it is appropriate to consider Wannier excitons, without strong vibrational dressing.
    In contrast, in this paper we consider organic molecules, and thus a Frenkel exciton picture, with strong vibrational dressing, and incoherent hopping is the principal mechanism of charge transport.
    Understanding how incoherent charge transport is modified by polariton condensation is a key ingredient toward realizing electrically driven organic polariton condensates.

    The questions of modification of chemical reaction rates and of charge transport are closely related, since many chemical reactions can be understood as electron transfer processes.  
    This is particularly true for non-adiabatic chemical reactions, where reaction rates are determined by Fermi-Golden rule transition rates between reactant and product potential energy surfaces~\cite{MartinezMartinez2018a,mandal2021theory,li2021cavity,schafer2021shining,Yang2021,li2021collective}. Thus this is a similar calculation to incoherent hopping transport~\cite{HerreraPRL16,Schafer2019}.
    Polariton condensation however changes material properties in additional ways, so the physics we discuss in this paper goes beyond calculations of incoherent hopping transport ``in the dark''.  
    The question of how chemical reactions are affected by a potential condensate of vibrational polaritons---resulting from strong coupling between vibrational modes and infrared photons---has been recently considered~\cite{Pannir22}, providing a complementary example of this point.

    In this paper we explore two main questions.  How electronic hopping can induce transitions between states (through exciting polaritons, or vibrational sidebands), and what determines the effective hopping rates to these states.  These questions are related, as the effective hopping rates require first identifying what final states can be reached, and summing over the rates of transitions to these individual states.   We find that a range of final states are possible.  The hopping matrix elements to some states (such as exciting the upper polariton) are suppressed in the thermodynamic limit (where there are many molecules), but a range of possible final states still exist: excitations of ``dark'' exciton states, and vibrational sidebands of the lower polariton.  However, at typical temperatures the dominant process is that leaving the system with the same macroscopically occupied lower polariton state.  This picture then allows a simplified calculation of how hopping rate varies with matter-light coupling, exciton-photon detuning, and polariton excitation density.
    
    The remainder of this paper is arranged as follows.  In Sec.~\ref{sec:model} we introduce the model we use to describe the molecular states, and the form of hopping operator that describes transitions between these states. To separate effects of exciton delocalization from those of polaron formation, Sec.~\ref{sec:hop-novib} discusses the case where we neglect coupling to vibrational states.
    We then extend this by including vibrational modes, and thus vibrational sidebands in Sec.~\ref{sec:hop-vib}.  In Sec.~\ref{sec:abs} we also discuss why vibrational sidebands do exist in hopping, but not in optical absorption.  Having established the dominant final state, section~\ref{sec:controlling} discusses how the hopping rate depends on matter-light coupling. Appendices provide details of the numerical method used throughout the paper, based on permutation symmetry~\cite{zeb2017,Zeb2022:FockMap}, as well as further numerical results provided for completeness.

    \section{Model}
    \label{sec:model}
    
    \subsection{Holstein--Tavis--Cummings and Holstein models}
    
    \begin{figure}[htpb]
      \centering
     \includegraphics[width=1\columnwidth]{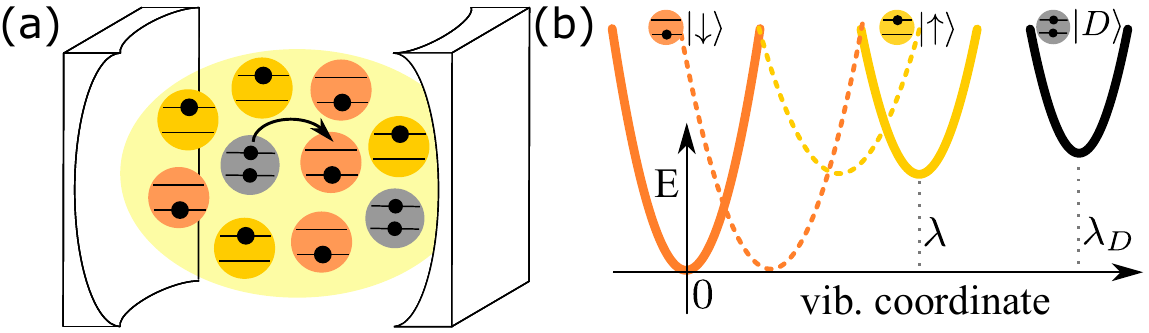}
      \caption{ 
      Sketch of the system.
      (a) 
      Organic molecules in an optical cavity.      
      The molecules are shown as two levels, HOMO and LUMO that can be empty or occupied with electrons (small dots). 
      Charge hopping (black arrow) occurs between charged (doubly occupied) and neutral (singly occupied) molecules. 
      The strong coupling between the neutral molecules and the cavity produces polaritons that can form a condensate and thus alter the charge hopping rates.
      (b)
      Potential energy surfaces corresponding to different electronic states. The dotted lines indicate the effective potential energy surfaces of the neutral molecules that form the polariton condensate (see Sec.~\ref{sec:controlling}).
     }  \label{fig:dhop}
    \end{figure}
    
    We describe the electronic state of molecules through two electronic levels, the highest occupied molecular orbital (HOMO) and the lowest unoccupied molecular orbital (LUMO), see Fig.~\ref{fig:dhop}.
    For simplicity, we assume identical  molecules, and neglect electron spin.  
    For such a model, four electronic states exist: Two neutral states, with a single electron in the HOMO ($\kdn$) or LUMO ($\kup$) levels, and two charged states, a positive empty molecule ($\ket{0}$), or a negative doubly occupied molecule ($\ket{D}$).
    The molecules are placed in an optical cavity, described as a single optical mode, which couples to transitions between the $\kdn$ and $\kup$ states, as in the Tavis--Cummings model~\cite{tavis-cummings1,tavis-cummings2}.  The cavity does not interact with the charged states.

    To model vibrational dressing of the electronic states, we include a single intramolecular vibrational mode.   For the optically active molecules we thus have the widely-used Holstein--Tavis--Cummings (HTC) model~\cite{Cwik2014,zeb2017}:
    \begin{multline}
      \label{eq:htc}
      H^{HTC}=
      \omega_c \ad \an
      +\sum_{n\in\aset} 
       \bigg\{
         \omega_0 \hat \sigma^+_n \hat \sigma^- _n+ 
         \frac{\wrr}{\sqrt{N}}  \left( \hat \sigma^+_n \hat a^{} + \hat \sigma^-_n \hat a^\dagger \right)
         \\
      +
     \wv 
        \left[
          \hat b^\dagger_n \hat b^{}_n
          - 
          \lam \hat\sigma^+_n \hat \sigma^-_n 
          ( \hat b^\dagger_n + \hat b^{}_n ) 
          \right] \bigg\}.
    \end{multline}
    Here $\an$ describes cavity photons with energy $\omega_c$, while the $N$ optically active molecules are described by Pauli operators $\hat \sigma_n$, acting in the $\ket{\ua},\ket{\da}$ subspace, with energy splitting $\omega_0$.  We denote the set of such optically active molecules as $\aset$.
    The collective Rabi splitting $\wrr$ parameterizes the matter-light coupling.
    As we make a rotating wave approximation, the number of excitations
    $N_{ex} = \ad \an + \sum_{n\in\aset} \hat \sigma^+_n \hat \sigma^-_n$ is conserved.

    The operator $\bn_n$ describes a vibrational mode with energy $\wv$,
    and vibrational coupling $\lam$.   We measure vibrational displacement with reference to the equilibrium for the $\kdn$ state.  As such,
    $\lam$ indicates the offset between the optimal vibrational displacement for the $\kup$ and $\kdn$ states.
     In reality, organic molecules have many vibrational and rotational modes, and different electronic states  displace different patterns of these. Our model relies on the common observation that a small number of modes dominate coupling to the electronic state.

    For the negatively charged molecules, there is no coupling to light, so each such molecule evolves independently.  For one such molecule we have the simpler Holstein model~\cite{Holstein1,Holstein2},
    \begin{eqnarray}
    \label{eq:hp}
       H^{H} =  \ket{D}\bra{D}\left\{ \wh + \wv\left[
          \hat b^\dagger\hat  b
          - 
          \lamp( \hat b^\dagger + \hat b ) \right] \right\},
    \end{eqnarray}
    where $\wh$ is the bare energy of the doubly occupied state, $\hat b$  the \emph{same} molecular vibrational mode as considered in Eq.~\eqref{eq:htc}.
    The parameter  $\lamp$ indicates the offset between the optimal vibrational displacement for the $\kD$ and $\kdn$ states.
    This Hamiltonian can be diagonalized by the Lang--Firsov (polaron) transformation~\cite{Lang1963a,Lang1963b}:
     \begin{align}
         U_{LF} &=\exp\left[
         \lambda_D\ket{D}\bra{D}(\hat{b}-\hat{b^\dagger})
         \right]\nonumber
         \\
         U_{LF} H^{H} U^\dagger_{LF} &= \ket{D}\bra{D}\left\{ \wh -\lamp^2\wv + \wv
          \hat b^\dagger\hat  b\right\}.
     \end{align}

    In the following we will use $\ket{\Psi^j}$ to denote the $j$th eigenstate of the neutral molecules, Eq.~(\ref{eq:htc}). For the charged molecule(s), we denote the $k$th such resulting eigenstate as $\ket{\Phi^k}$.

    \subsection{Hopping processes}
    
    Incoherent charge transfer between neighboring molecules can occur due to tunneling matrix elements.
    Hopping can proceed via two channels, LUMO-LUMO (labeled $L$) which interchanges molecules in the states $\ket{D},\ket{\da}$
    and HOMO-HOMO (labeled $H$)
    which interchanges molecules in the states $\ket{D},\ket{\ua}$.
    Figure~\ref{fig:dhop}(a) illustrates hopping in the $L$ channel.
    We will consider a single hopping process at a time. As such, we will consider  a \emph{single} negatively charged molecule described by Eq.~\eqref{eq:hp} along with $N$ neutral optically active  molecules described by Eq.~\eqref{eq:htc}. 
    The operators describing hopping from molecule $p$ to $q$ are
    \begin{equation}
        \hat V_{pq}^{L}=  \ket{\da_{p}D_q}\bra{D_p\da_q}
        , \quad
        \hat V_{pq}^{H} =  \ket{\ua_{p}D_q}\bra{D_p\ua_q}.   
    \end{equation}
    Associated with these operators are bare hopping amplitudes $J^L, J^H$  in each channel.
    If we were to consider positively charged molecules and hopping of holes, the relation between $L$, $H$ and $\da$, $\ua$ would swap.

    Below, we will calculate the probabilities and energies of the final states after  hopping, and thus find how the overall hopping rate is modified by the presence of a polariton condensate.
     Before hopping, we assume the whole system is in the lowest energy state for a given number of excitations $N_{ex}$.  This state is a condensate of lower polaritons along with a charged molecule, $p$, in its relaxed state. Using the notation for eigenstates introduced above,  this state can be written as $\ket{\Psi_{\aset^\prime \cup \{q\}}^{0}\Phi_{p}^{0}}$.
    Here $\aset^\prime$ indicates the set of  $N-1$ molecules in the active sector not involved in the hopping, while $\aset^\prime \cup \{q\}$ indicates the set of all active molecules before the hopping event, which includes also the molecule $q$ that is involved in the hopping. 
    In the following we will take the energy of this state as a reference $\varepsilon_{0,0}\equiv 0$.
    
    After the hopping process, the set of active molecules will become
    $\aset^\prime \cup \{p\}$, and molecule $q$ will be charged.
    As well as changing which molecule is charged,  hopping can cause transitions to  excited states, $\ket{\Psi_{\aset^\prime \cup \{p\}}^{j}\Phi_{q}^{k}}$, at energies $\varepsilon_{j,k} \geq 0$.
    With this notation, we can define hopping matrix elements
    \begin{equation}
       \widetilde{\matelem}^{l_{j,k},\text{tot}.}\equiv \left|\mel{\Psi_{\aset^\prime\cup \{p\}}^{j}\Phi_{q}^{k} }{ \displaystyle\sum_{c=L,H} J^c \hat V_{pq}^c }{\Psi_{\aset^\prime \cup \{q\}}^{0}\Phi_{p}^{0} }\right|^2,
       \label{eq:defMtot}
    \end{equation}
    where ${c\in\{L,H\}}$ denotes the hopping channels, and $l_{j,k}$ indexes the final state.  
    In cases where $J^L \gg J^H$ or vice-versa, hopping will be dominated by a single channel, and we may consider the single channel matrix elements
    \begin{equation}
       \matelem^{l_{j,k}(c)}\equiv \left|\mel{\Psi_{\aset^\prime\cup \{p\}}^{j}\Phi_{q}^{k} }{\hat V_{pq}^c }{\Psi_{\aset^\prime \cup \{q\}}^{0}\Phi_{p}^{0} }\right|^2.
       \label{eq:defM}
    \end{equation}
    When $J^{L,H}$ are comparable interference between the two hopping channels can occur.

    Transitions to excited states are possible because the separate channel hopping processes effectively measure the electronic state of the hopping molecule.  Restricting to the active molecule involved in the hopping, and ignoring the fact its location changes, the hopping processes have the effect $\hat V^{L}_p= \hat \sigma^-_p \hat \sigma^+_p$ and $\hat V^{H}_p= \hat \sigma^+_p \hat \sigma^-_p$, where we have used
    $p$ to denote the molecule $q/p$ before/after hopping.
    That is, hopping in the LUMO channel requires an active molecule in the $\da$ state, while hopping in the HOMO channel requires an active $\ua$ state.
    As such, by using completeness of the final states, we see that the matrix elements in a given channel sum to give 
    \begin{equation}
        \label{eq:sumrule}
        \sum_{j,k} \matelem^{l_{j,k}(c)}= p_{\sigma(c)},
    \end{equation}
    where $p_{\sigma}$ is the probability to find the active molecule in the $\ket{\sigma}$ state, with $\sigma(L)=\da$, $\sigma(H)=\ua$.
    By measuring the state on a single molecule, these operations can mix different polaritonic eigenstates. 
    We may also note that $\hat V^L + \hat V^H=\mathbb{1}$ within the electronic sector.  That means that in the special case $J^L=J^H$, interference between the channels prevents the electronic state changing.  The vibrational state may though still change.
    It also means that (neglecting vibrations) when $j,k \neq 0,0$, one has that the single-channel matrix elements are independent of channel,
    $\matelem^{l_{j,k}(H)}=\matelem^{l_{j,k}(L)}$.
    
    Since transitions to states with $\varepsilon_{j,k} \geq 0$ describe an increase in energy of the molecular system, they require extracting energy from a thermal reservoir---either delocalized phonon modes, or low energy intramolecular vibrational modes not explicitly included in our model.  
    This energy cost leads to Boltzmann weights for excited state processes, giving an overall hopping rate~\cite{SchmidlinPMB80,Bassler99a,Bassler99b,Pope}:
    \begin{equation}
      \label{eq:hoprate}
      R =  \sum_{j,k} \widetilde{\matelem}^{l_{j,k}, \text{tot}.} e^{-\varepsilon_{j,k}/k_BT}.
    \end{equation} 
    The charge mobility is  proportional to the hopping rate $R$~\cite{SchmidlinPMB80,Bassler99a,Bassler99b,Pope}.  In the following we will discuss how to evaluate $\matelem^{l_{j,k}(c)}$ in various cases, and thus determine hopping rates.

    \section{Hopping-induced transitions neglecting vibrations}
    \label{sec:hop-novib}
    
    In this section we look at hopping without vibrational modes.  This is equivalent to setting $\lam=\lamp=0$, so that all molecules remain in the vibrational ground state and Eq.~\eqref{eq:htc} becomes the Tavis--Cummings  model~\cite{tavis-cummings1,tavis-cummings2}.
    We do this to enable us to understand separately the effects of exciton delocalization (present in this section) and those of polaron formation (present in later sections with vibrations).
    
    Without vibrational dressing, the charged molecule has only a single state $\ket{D}$. 
    As such, the states before and after hopping can be written as $\ket{\Psi_{\aset^\prime \cup \{q\}}^{0}D^{}_p}$ and $\ket{\Psi_{\aset^\prime \cup \{p\}}^{j}D^{}_q}$, and a single index $j$ identifies the final state. 
    To enumerate the final states of the active sector, we must consider eigenstates of the  Tavis--Cummings model.  These are formed of three kinds of excitations:  lower polaritons (LP), upper polaritons (UP), and dark states.  
    Polariton states involve superpositions of photons and uniformly delocalized matter excitations, as created by the operator $\sum_{n \in \aset} \hat{\sigma}^+_n/\sqrt{N}$.  The dark states correspond to the $N-1$ degenerate modes of matter excitons which are orthogonal to this uniform mode.  One possible basis for dark states is the Fourier basis $\sum_{n \in \aset} e^{i 2\pi k n/N} \hat{\sigma}^+_n/\sqrt{N}$ for $k=1 \ldots N-1$. 
    However, since dark states are degenerate, any basis spanning this space is suitable.
    In writing this expression for dark states we have implicitly assumed the sites $n\in \aset$ can be numbered $n=1\ldots N$; we will continue to assume this in the remainder of this article.

    In the following we will first discuss in Sec.~\ref{sec:novib-small-rho} the simple picture that occurs when $N_{ex} \ll N$, where analytic results are possible. 
    Section~\ref{sec:novib-arbit-rho} then presents numerical results at arbitrary excitation density $\rho_{ex}=N_{ex}/N$.  We conclude this vibration-free discussion with analytic results in the other extreme limit, where $N_{ex} \gg N$, given in Sec.~\ref{sec:novib-large-rho}.

    \subsection{Analytic matrix elements at small excitation density}
    \label{sec:novib-small-rho}
    
    In the limit where $N_{ex} \ll N$, the many particle states take a simple form.  To see this, we start by defining operators:
    \begin{align}
        \hat{c}^\dagger_{LP}
        &=
        \cos\theta \hat{a}^\dagger
        -
        \frac{\sin\theta}{\sqrt{N}}
        \sum_{n \in \aset}  \hat{\sigma}^+_n,
        \\
        \hat{c}^\dagger_{UP}
        &=
        \sin\theta \hat{a}^\dagger
        +
        \frac{\cos\theta}{\sqrt{N}}
        \sum_{n \in \aset}  \hat{\sigma}^+_n,
        \\
        \hat{d}^\dagger_k 
        &= 
        \frac{1}{\sqrt{N}}
        \sum_{n \in \aset} e^{i 2\pi k n/N} \hat{\sigma}^+_n,
    \end{align}
    where $\theta$ is the Hopfield angle, $\tan(2\theta)=2 \wrr/(\omega_0-\omega_c)$.  When $N_{ex}\ll N$, these operators approximately obey bosonic commutation relations, and the system eigenstates are approximately given by number states (Fock states) of these operators. 
    At higher density---as is discussed in subsequent sections---the states are modified because of saturation of the two-level systems.
    
    Since hopping changes the state of only one molecule, there are restrictions on the final states that can be reached in this low excitation limit.   In the low excitation limit, one can invert the definitions of $\hat{c}^\dagger_{LP,UP}$, $\hat{d}^\dagger_k$ to write $\hat\sigma^+_{p}$ as a linear combination of these operators.
    As such, the hopping operators $\hat\sigma^+_p\hat\sigma^-_p$ and $\hat\sigma^-_p\hat\sigma^+_p$ correspond to a quadratic operation which can scatter at most one particle to the UP and dark modes.  That is, the possible final states involve $N_{ex}-1$ lower polaritons, and one excitation which is in either the LP, UP, or a dark state.   As we will discuss below, while this statement is only strictly true for $N_{ex} \ll N$, it can be shown to be approximately true much more broadly, as long as $N \gg 1$.

    \subsubsection{\texorpdfstring{$N_{ex}=1$}{Nex=1} case}
    For ${N_{ex}=1}$, the probabilities have  closed forms, which also help explain behavior at $N_{ex}>1$.
    At resonance, i.e.,~$\omega_c=\omega_0$, the $N_{ex}=1$ LP and UP states are:
    \begin{equation}
    \ket{\Psi^{LP/UP}}=\frac{1}{\sqrt{2}}\left[\frac{1}{\sqrt{N}}\sum_{n=1}^N\ket{0_P;\uparrow_{n}\Downarrow_{\neq n}}
    \mp \ket{1_P;\Downarrow}\right],
    \end{equation} 
    where $\ket{0_P}, \ket{1_P}$ denote the photon states with $0$ or $1$ photons, and $\ket{\uparrow_{n}\Downarrow_{\neq n}}$ indicates the molecular electronic state where the $n$th molecule is excited and all other molecules unexcited.
    The ${N-1}$ dark exciton states can be written as:
    \begin{equation}
      \ket{\Psi^{d_k}} = \sum_{n=1}^N\frac{e^{i2\pi k n/N}}{\sqrt{N}}
      \ket{0_P;\uparrow_{n}\Downarrow_{\neq n}},
      \;\;\;
      k\in[1,N-1].
    \end{equation}
    In this notation the state before hopping is $\ket{\Psi^{LP}_{\aset^\prime \cup \{q\}}D_p}$.
    To find the probabilities for the $L$ or $H$ channel, we  project onto the space where molecule $q$ is in the $\da$ or $\ua$ state respectively.
    This yields ${\matelem^{LP(H)}= 1/4N^2}$,
    ${\matelem^{LP(L)}=(1 - 1/2N)^2}$.
    
    For other final states, we use the result noted above that for $j\neq LP$, the matrix element $\matelem^{j(c)}$ is independent of channel label $c\in\{L,H\}$.
    For the UP we find $\matelem^{UP(c)}=1/4N^2$, while
    for  dark states as defined above we have  ${\matelem^{d_k (c)}=1/2N^2}$ independent of $k$.
    Summing over all dark states gives a total probability ${\matelem^{\text{Dark} (c)}= (N-1)\matelem^{d_k (c)}=(N-1)/2N^2}$.

    One may note that for this resonant case at large $N$, transitions to dark states saturates the sum rule for the HOMO channel, $ \sum_{j} \matelem^{{j}(H)}=1/(2N)$.
    In contrast, for the LUMO channel, the sum rule  $ \sum_{j} \matelem^{{j}(L)}=1$ is saturated by the transition to the lower polariton state. Thus, for $N_{ex}=1$, in the limit $N\to \infty$, the only surviving process is a transition to the LP through the LUMO channel. This occurs because exciton delocalization means local hopping only perturbs the state by an amount $\propto 1/\sqrt{N}$.

    \subsubsection{\texorpdfstring{$N_{ex}=2$}{Nex=2} case}
    
    Closed forms can also be found for $N_{ex}=2$, which allow one to understand why the probability to create multiple excitations remains small at arbitrary $N_{ex}/N$, even though such processes are not forbidden.
    
    Considering first the polaritonic states, these are formed from a basis of photon and bright excitonic states which we write as:
    \begin{align*}
        \ket{1_P;B}
        &=
        \frac{1}{\sqrt{N}}
        \sum_{n=1}^N\ket{1_P;\uparrow_{n}\Downarrow_{\neq n}},
        \\
        \ket{0_P;BB}
        &=
        \frac{1}{\sqrt{2N(N-1)}}
        \sum_{\substack{n,m=1\\n\neq m}}^N
        \ket{0_P;\uparrow_{n}\uparrow_{m}\Downarrow_{\neq n,m}},
    \end{align*}
    along with the two photon state $\ket{2_P;\Downarrow}$.  Writing the Tavis--Cummings Hamiltonian in the basis
    $\ket{2_P;\Downarrow},\ket{1_P;B},\ket{0_P;BB}$ one finds that at resonance, $\omega_c=\omega_0$, the eigenstates are
    \begin{align}
    \ket{\Psi^{2LP/2UP}}&=
    \frac{1}{\sqrt{2(1+\eta^2)}}
    \begin{pmatrix}
    1 \\ \mp\sqrt{1+\eta^2} \\ \eta
    \end{pmatrix},
    \\
    \ket{\Psi^{LP+UP}}&=
    \frac{1}{\sqrt{1+\eta^2}}
    \begin{pmatrix}
    \eta \\ 0 \\ -1
    \end{pmatrix}, \quad
    \eta=\sqrt{\frac{N-1}{N}}.
    \end{align}
    Projected into this same basis, the HOMO hopping operator $\hat V^{H}$ is a diagonal matrix (since it cannot change photon number) with diagonal elements $(0,1,2)/N$.  This gives
    \begin{equation}
        \matelem^{j(H)}=
        \frac{1}{[N(4N-2)]^2}
        \begin{cases}
        (4N-3)^2 & j=2LP \\
        8N(N-1) & j=LP+UP \\
        1 & j=2 UP
        \end{cases}.
    \end{equation}
    Notably while the first two terms here are $\mathcal{O}(N^{-2})$, the last is $\mathcal{O}(N^{-4})$, consistent with the suppression of transitions changing multiple excitations.  For hopping in the LUMO channel, as discussed above we have $\matelem^{j(H)}=\matelem^{j(L)}$,  except for $j=2LP$.  For that case we get
    \begin{equation}
        \matelem^{2LP(L)}=
        \left[1-
        \frac{(4N-3)}{N(4N-2)}
        \right]^2.
    \end{equation}
    This saturates the LUMO channel sum rule at large $N$, and we again find that LUMO channel hopping with the state unchanged is the only term that survives in the $N\to \infty$ limit.
    
    While the above shows individual matrix elements for final states differing by more than one excitation are suppressed, one may note that considering the dark excitonic states, there are $\mathcal{O}(N^2)$ states with two dark excitons, compared to $\mathcal{O}(N)$ with one.  As we next show, despite this counting effect, the total weight of transitions to the sector with two dark excitons is suppressed by $1/N$.
    
    The dark exciton states can be written as
    \begin{equation}
        \ket{0_P;d_k d_{k^\prime}}
        =
        \sum_{\substack{n,m=1\\n\neq m}}^N
        \frac{e^{i 2\pi(k n + k^\prime m)/N}}{\sqrt{N(N-2)}}
        \ket{0_P;\uparrow_{n}\uparrow_{m}\Downarrow_{\neq n,m}},
    \end{equation}
    as long as $k \neq k^\prime$. (When $k=k^\prime$, the normalization of this state changes.  Since $k\neq k^\prime$ makes the dominant contribution to the sum over dark states, we focus only on this case for simplicity). 
    The state with a single dark exciton and one bright exciton is a special case of this, $\ket{0_P;d_k B}=\ket{0_P;d_k d_0}$.   One may show that
    \begin{align}
        \mel{0_P;d_k B}{\hat V^H_p}{0_P;BB}
        &=
        \frac{e^{i 2\pi k p/N}}{N}
        \sqrt{\frac{2 (N-2)}{(N-1)}},
        \\
        \mel{0_P;d_k d_{k^\prime}}{\hat V^H_p}{0_P;BB}
        &=
        \frac{-\sqrt{2}e^{i 2 \pi (k+k^\prime)p/N}}{N\sqrt{(N-1)(N-2)}}.
    \end{align}
    Without further calculation, one may see that after squaring these rates and summing over the number of final states, the total rate of transitions to states with one dark exciton will be $\mathcal{O}(N \times N^{-2})$ while transitions to states with two dark excitons are $\mathcal{O}(N^2 \times N^{-4})$.  Thus, transitions to states with multiple dark excitons are indeed suppressed.
    Moreover, by constructing the eigenstate
    $\ket{\Psi^{LP+d_k}}=(\ket{0_P;d_k B}-\ket{1_P;d_k})/\sqrt{2}$ and using results for matrix elements in the one and two excitation subspaces one finds
    \begin{equation}
        \matelem^{LP+d_k(c)}
        = \frac{1}{4 N^2}
        \left[
        \sqrt{\frac{2N-4}{2N-1}}+1
        \right]^2.
    \end{equation}
    Note that we have again used that $\matelem^{j(c)}$ is independent of $c$ when $j \neq 2LP$.
    Summing over the dark states, 
    $\matelem^{\text{LP+Dark}(c)}\equiv(N-1)\matelem^{LP+d_k(c)}=1/N + \mathcal{O}(1/N^2)$ and so
    one again finds these processes saturate the sum rule for $N_{ex}=2$ states,
    $\sum_{j} \matelem^{l_{j}(H)}=1/N$.

    \subsection{Numerical matrix elements at arbitrary excitation density}
    \label{sec:novib-arbit-rho}

     We next consider behavior at finite $\rho_{ex}\equiv N_{ex}/N$. Brute force calculations here are challenging, as the Hilbert space of the Tavis--Cummings model scales exponentially with $N$.  Fortunately, for identical molecules, we can exploit permutation symmetry to reduce the scaling  to $\mathcal{O}(N)$, which enables calculations even at $N\sim10^3$, see Ref.~\cite{zeb2017,Zeb2022:FockMap} and Appendix~\ref{sec:permutations} for details.  Note that in doing this we must treat the molecule involved in the hopping separately from the others.

    \begin{figure}[htpb]
      \centering
    \includegraphics[width=1\columnwidth]{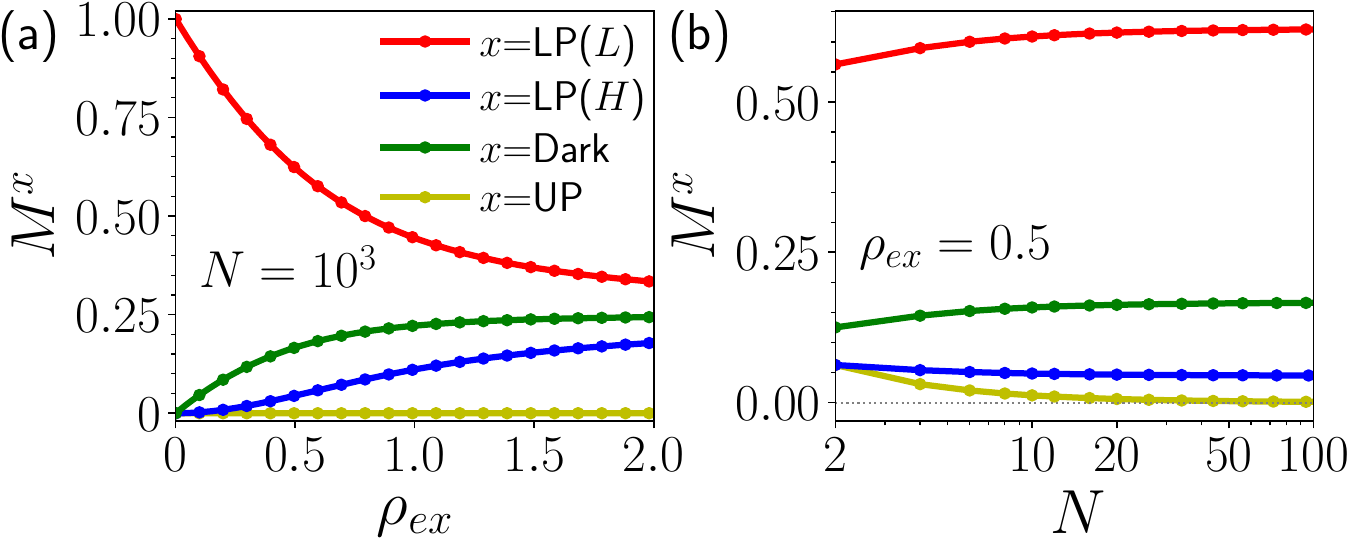}
      \caption{
        Probabilities for hopping to produce a given state, neglecting vibrations, via channels $L,H$.  For dark and UP final states, the result is independent of channel.  
        (a) vs excitation density $\rho_{ex}$ at $N=10^3$. (b) vs number of molecules $N$ at  $\rho_{ex}=0.5$. Plotted on resonance, $\omega_c=\omega_0$; in this limit the figure is independent of the value of $\wrr$.}
      \label{fig:novib-prob}
    \end{figure}

    Figure~\ref{fig:novib-prob}
    shows the behavior of the matrix elements as a function of excitation density $\rho_{ex}$ at fixed $N$, and vs $N$ at fixed $\rho_{ex}$.
    Since the only final states with significant weight are those with one excitation, we will abbreviate the matrix element $\matelem^{(N_{ex}-1)LP + x (c)}$ as
    $\matelem^{x(c)}$, where
    $x \in \{LP, UP, \text{Dark}\}$.
    Figure~\ref{fig:novib-prob}(a) shows that at small $\rho_{ex}$, the state-changing probabilities grow linearly with $\rho_{ex}$, so $\matelem^{UP(c)}\simeq N_{ex}/4N^2$ and  $\matelem^{\text{Dark}(c)}\simeq N_{ex}/2N$.
     Increasing $\rho_{ex}$ equalizes the probability of finding a given molecule excited or unexcited.  As such, at large $\rho_{ex}$, one finds $\matelem^{LP(L)}$ decreases and $\matelem^{LP(H)}$ increases, with both elements approaching $1/4$ at large $\rho_{ex}$.
    In this same limit,  the probability $\matelem^{UP(c)}$ vanishes as $1/4N$.
    On the other hand, $\matelem^{\text{Dark}(c)}$  saturates at $1/4$, matching the $LP$ state.  These results match analytic results available at large excitation density, discussed in the next section.
    Figure~\ref{fig:novib-prob}(b) shows the $N$-dependence at intermediate $\rho_{ex}$, showing which terms vanish or remain finite in the large $N$ limit.

    \begin{figure}[htpb]
        \centering
        \includegraphics[width=1\columnwidth]{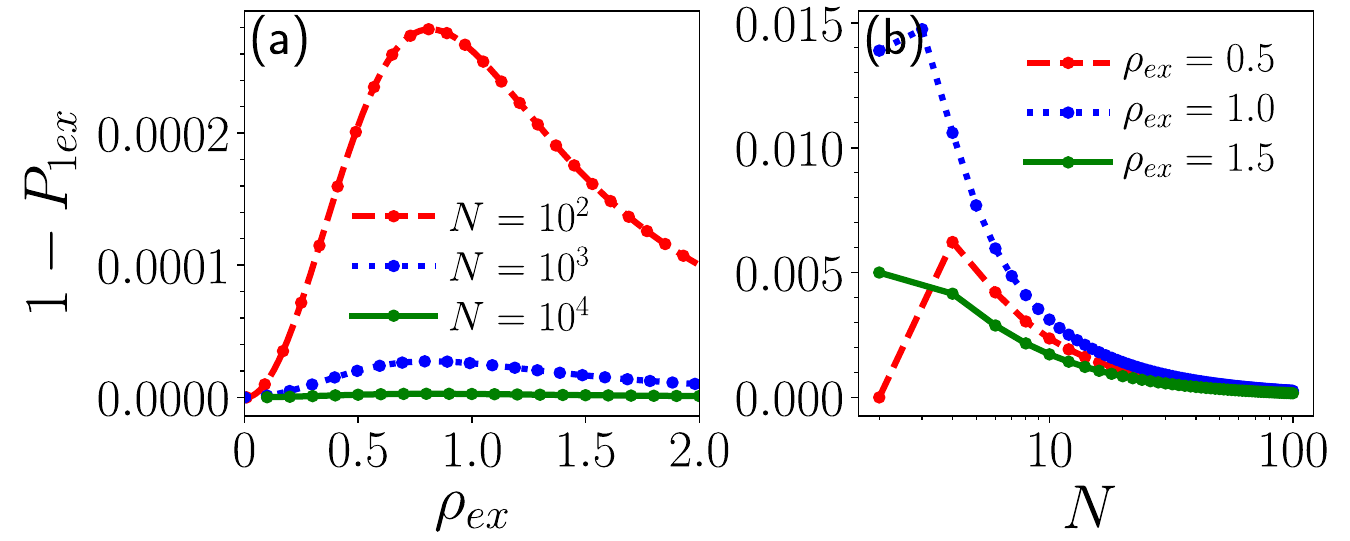}
        \caption{Probability to reach a final state differing by more than one excitation from the initial state (see Eq.~\eqref{eq:P1ex}).  (a) Probability vs excitation density at fixed values of $N$ as indicated.  (b) Probability vs $N$ at fixed $\rho_{ex}$ as indicated.  Plotted on resonance, $\omega_c=\omega_0$, and thus the figure is independent of $\wrr$.
        Note that for $\rho_{ex}=0.5$, the left-most point in panel (b) corresponds to $N=2, N_{ex}=1$, thus final states with two excitations are not possible.
        }
        \label{fig:novib-other-states}
    \end{figure}
    
   As noted above, while transitions to states with multiple excitations are possible, their weight is suppressed at large $N$.  Figure~\ref{fig:novib-other-states} shows  numerically that this remains true even for non-vanishing $N_{\text{ex}}/N$.  Specifically, defining 
   \begin{equation}
       \label{eq:P1ex}
       P_{1ex}=\sum_{x \in LP, UP, \text{Dark}} \sum_c \matelem^{(N_{ex}-1)LP+x(c)},
   \end{equation}
   then any deviation of $P_{1ex}$ from 1 indicates the total amplitude of processes producing multiple excitations, which is seen to be small.

    Although the probabilities for hopping to excite dark states grow with $\rho_{ex}$, the dominant process in the hopping rate $R$ remains the LP channel at all relevant temperatures.
    This is because the Boltzmann weights in Eq.~\eqref{eq:hoprate} suppress excited final states, so LP state dominates the hopping rates if $k_B T \ll \wrr$.

    \subsection{Analytic matrix elements at large excitation density}
    \label{sec:novib-large-rho}
    
    Analytic results for hopping matrix elements can also be found in the limit where $N_{ex} \gg N$. These help explain the numerical results found at general $N_{ex}/N$.
    
    To find the ground state in the limit $N_{ex} \gg N$, we may note that in this limit the photon mode will always be highly occupied.  Furthermore, the matrix element for photon raising and lowering operators between sequential number states will always be approximately $\sqrt{N_{ex}}$, as the difference between states with $N_{ex}$ and $N_{ex}-N$ photons can be neglected.
    If we choose a state where alternating photon number states have opposite signs, this means that the Tavis--Cummings Hamiltonian becomes $H_{TC} \simeq - \wrr \sqrt{N_{ex}/N} \hat S^x$, where $\hat S^x = \sum_n (\hat\sigma_n^++\hat\sigma_n^-)/2$ is a collective spin operator.
    The ground state of $H_{TC}$ in this limit is a state with collective spin aligned along the $x$ axis; this is equivalent to
    $\sum_{\{\sigma\}}\ket{\{\sigma\}}/\sqrt{2^N}$ where we have used $\{\sigma\}$ to denote summation over all  configurations of the spin states in the $\hat \sigma^z$ basis, $\sigma_n\in \{\uparrow, \downarrow\}$.
    As a result, the ground state at $N_{ex} \gg N$---i.e.~the state corresponding to $(N_{ex}-1)$ LP excitations---can be approximated by:
    \begin{equation}
      \label{eq:largerhogs}
      \ket{\Psi^0(N_{ex})} \simeq 
      \sum_{\{\sigma\}} 
      \frac{(-1)^{\CountSigma_{\{\sigma\}}}}{\sqrt{2^N}}
      \ket{(N_{ex}-\CountSigma_{\{\sigma\}})_P;\{\sigma\} },
    \end{equation}
    where $\CountSigma_{\{\sigma\}} =   \mel{\{\sigma\}}{\sum_n \hat \sigma^+_n \hat\sigma^-_n}{\{\sigma\}}$  counts the excited molecules.
    As previously $\ket{(m)_P}$ denotes the photon number state $m$.  
    The state in Eq.~\eqref{eq:largerhogs} thus takes the equally weighted spin configuration, and adjusts the photon numbers to fix the  total excitation number.  The signs ensure the photon matrix elements have negative signs. 
    Since all spin configurations have equal weight,  the expression $\mel{\Psi^0}{\hat V^{c}_p}{\Psi^0}=1/2$ corresponds to the fraction of terms where molecule $p$ is unexcited/excited respectively. As such, the channel-dependent transition probabilities of going to the unexcited final state,  $\matelem^{LP(c)}$ becomes $1/4$ for both values of $c$, as seen in Fig.~\ref{fig:novib-prob}(a).  
    
    Using the above state, we can also find the probabilities
    for transitions to states with a single dark exciton or upper polariton excited,
    $\matelem^{\text{Dark}(c)}$ and $\matelem^{UP(c)}$.  As noted above, in the absence of vibrations, both these amplitudes are independent of the channel label, as
    $\hat V^L + \hat V^H=\mathbb{1}$ in the relevant subspace for hopping. 
    
    We first consider the amplitude for dark states.  We must first find the large excitation density limit of the state $(N_{ex}-1)LP+d_k$  which, for brevity, we denote $\ket{\Psi^{d_k}}$.  Making use
    of Eq.~\eqref{eq:largerhogs} this can be written  as:
    \begin{displaymath}
      \ket{\Psi^{d_k}} \propto
      \frac{1}{\sqrt{N}}\sum_{n=1}^N e^{i 2\pi k n/N} \hat\sigma^+_n
      \ket{\Psi^0(N_{ex}-1)}.
    \end{displaymath}
    Clearly this involves $N_{ex}-1$ lower polaritons (as before), and one excitation in a finite $k$ state.  By considering the action of the spin
    raising operators we can rewrite this in a way that simplifies subsequent calculations:
    \begin{align}
      \label{eq:largerhodark}
      \ket{\Psi^{d_k}}
      &=
        \sum_{{\{\sigma\}}}
        \frac{(-1)^{\CountSigma_{\{\sigma\}}} \CountFTSigma_{k,\{\sigma\}} }{\sqrt{N 2^{N-2}}}
        \ket{(N_{ex}-\CountSigma_{\{\sigma\}})_P; \{\sigma\} },
      \\
      \CountFTSigma_{k,\{\sigma\}}
      &=
        \bra{\{\sigma\}}\sum_n e^{i 2\pi k n/N} \hat\sigma^+_n \hat\sigma^-_n\ket{\{\sigma\}}.
        \nonumber
    \end{align}
    The factor $\CountFTSigma_{k,\{\sigma\}}$ sums up the phase factors that could arise in producing a given final state. This form arises since exactly one of the excited spins must come from the dark state operator, so for each possible spin state, we must add a copy of the state with the corresponding phase factor.  To verify the normalization and work out the matrix elements it is useful to use the result
    \begin{align*}
    C_{n,n^\prime}&=
     \sum_{{\{\sigma\}}}
      \mel{\{\sigma\}}{%
      \left(\hat\sigma^-_n\hat\sigma^+_n\right)
      \left(\hat\sigma^-_{n^\prime}\hat\sigma^+_{n^\prime}\right)}{\{\sigma\}}
      \\&=2^{N-2}(1+\delta_{n,{n^\prime}}),
    \end{align*}
    from counting the number of spin configurations.
    The normalization factor can then be found using:
    \begin{displaymath}
      \sum_{{\{\sigma\}}}
      |\CountFTSigma_{k,\{\sigma\}}|^2
    = \sum_{n,n^\prime} e^{i 2 \pi k(n-n^\prime)/N} C_{n,n^\prime}
    = N 2^{N-2}.
    \end{displaymath}
    We can then find the relevant matrix elements:
    \begin{multline}
      \mel{\Psi^{d_k}}{\hat V^H_p}{\Psi^0} =
      \frac{1}{\sqrt{N 2^{2N-2}}}
      \sum_{{\{\sigma\}}}
      \CountFTSigma_{k,\{\sigma\}} \mel{\{\sigma\}}{\hat\sigma^+_p\hat\sigma^-_p}{\{\sigma\}}
     \\=\frac{1}{\sqrt{N 2^{2N-2}}}\sum_n e^{i 2 \pi k n/N} C_{p,n}
      = \frac{e^{i 2 \pi k p/N}}{2 \sqrt{N}}.
    \end{multline}
    Hence, summing over all dark states we find $\matelem^{\text{Dark}(c)} = (N-1)/4N$, matching Fig.~\ref{fig:novib-prob}.
    
    For transitions to the upper polariton---i.e. a state $(N_{ex}-1)LP + UP$---we can proceed in a similar way.
    To identify the state with exactly one upper polariton excitation, we note that this state should exist within the manifold described by the (symmetric) collective spin operators $\hat S^x$, $\hat S^y$,$\hat S^z$.  As such, we can consider the state with one upper polariton to be the first excited state in the symmetric sector.  This corresponds to acting once 
    on the state in Eq.~(\ref{eq:largerhogs})
    with the operator which lowers the collective $x$ spin by one unit. This operator is:
    $\sum_n (-\ket{\ua}_n + \ket{\da}_n)(\bra{\ua}_n+\bra{\da}_n)/2$.
    Ignoring photons, this state is thus:
    \begin{displaymath}
    \frac{1}{\sqrt{N}}
      \sum_n \bigotimes_m
      \left( \frac{(-1)^{\delta_{n,m}} \ket{\ua}_m + \ket{\da}_m}{\sqrt{2}} \right).
    \end{displaymath}
    Rewriting in terms of spin configurations as above, and re-introducing the photons and their sign factors gives
    \begin{multline}
      \label{eq:largerhoup}
      \ket{\Psi^{UP}(N_{ex})} \simeq \frac{1}{\sqrt{N2^N}}
      \sum_n \sum_{\{\sigma\}} (-1)^{\CountSigma_{n,\{\sigma\}}+\CountSigma_{\{\sigma\}}}
      \\\times
      \ket{(N_{ex}-\CountSigma_{\{\sigma\}})_P; \{\sigma\} },
    \end{multline}
    where $\CountSigma_{n,\{\sigma\}}=\mel{\{\sigma\}}{\hat\sigma^+_n\hat\sigma^-_n}{\{\sigma\}}$. One may easily check this state is normalized.   The overlap can then be found to
    be
    \begin{multline}
      \mel{\Psi^{UP}}{\hat V^H_p}{\Psi^0} =\\
      \sum_{{\{\sigma\}}}
      \sum_n 
      \frac{(-1)^{\CountSigma_{n,\{\sigma\}}}}{2^N\sqrt{N}}
       \mel{\{\sigma\}}{\hat\sigma^+_p\hat\sigma^-_p}{\{\sigma\}}
      =-\frac{1}{2\sqrt{N}},
    \end{multline}
    so 
    $\matelem^{\text{UP}(c)} = 1/4N$, consistent with the vanishing value at large $N$ seen in Fig.~\ref{fig:novib-prob}.
    
    
    \section{Hopping-induced transitions in the presence of vibrations}
    \label{sec:hop-vib}
    
    In the previous section we analyzed the behavior of hopping matrix elements in the Tavis--Cummings model, neglecting vibrational excitations.  In this section we explore a similar question regarding changes to the molecular vibrational state.  
    
    Due to the different vibrational offset of the electronic states $\kup,\kdn,\ket{D}$,  hopping can excite vibrational modes of both the charged and active molecules.
    In addressing this, one must note that the delocalized nature of polaritons alters the vibrationally dressed states. This question has been previously explored in the context of optical absorption~\cite{HerreraPRL16,Herrera2016a,wu2016,zeb2017}.  
    Comparing absorption to hopping processes poses an important question: is it possible to excite a vibronic sideband of the lower polariton condensate state?  
    For absorption, previous work~\cite{Cwik16,Herrera2016a,zeb2017} observed that there is only a single isolated lower polariton peak, with no vibronic sidebands.  
    As we show below, the situation differs for hopping.

    In the following we first discuss the excitations that can be created during hopping in Sec.~\ref{sec:vibexc}, and then discuss how this differs from those seen in the optical absorption spectrum in Sec.~\ref{sec:abs}.

    \subsection{Hopping response function}
    \label{sec:vibexc}

    \begin{figure}[htpb]
      \centering
      \includegraphics[width=1\columnwidth]{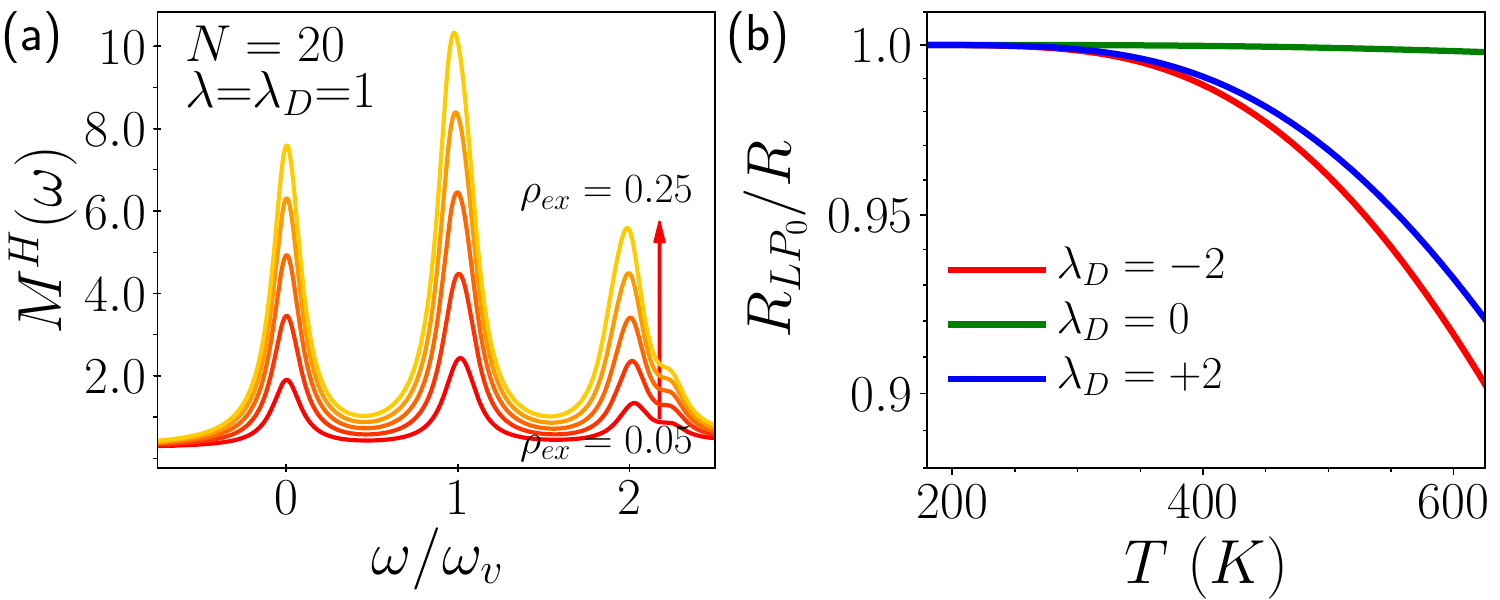}
      \caption{
      (a) 
      Vibrational sidebands near the LP state induced by electron hopping,  seen via the response function $\hopresp^{(H)}(\omega)$.  Lines correspond to
      varying  $\rho_{ex}$ from $0.05$ (bottom) to $0.25$ (top) in steps
      of $0.05$. Plotted for $N=20$, $\omega_0=\omega_c$, $\wrr=1$eV, $\lam=\lamp=1$, $\wv=0.2$eV.
      Frequencies are measured from the LP energy 
      and a linewidth of $0.02$eV is added to broaden the peaks.
      (b)
      Fraction of total hopping rate associated with the final state being the unexcited lower polariton ($LP_0$) vs temperature. This illustrates the effect of the Boltzmann weight of transitions to excited vibrational states.   Shown for three values of $\lamp$, with $\lam=1$.
      }
      \label{fig:response}
    \end{figure}

    To illustrate the potential excitations created by hopping, we consider a hopping response function, defined by analogy with the optical response function (see below):
    \begin{align}
        \label{eq:hoppingresp}
        \hopresp^{(c)}(t)
        &\equiv \sum_{j,k} \matelem^{l_{j,k}(c)} e^{-i\varepsilon_{j,k}t}
        \nonumber\\&=
        \mel{\Psi_{\aset^\prime \cup \{q\}}^0\Phi_p^0 }{\hat V_{qp}^c(t)\hat V_{pq}^c(0)}{\Psi_{\aset^\prime \cup \{q\}}^0\Phi_p^0}.
    \end{align}
    By defining this function in the time domain, it allows straightforward calculation using the permutation symmetric basis approach, see Appendix~\ref{sec:permutations}, and in particular
    Sec.~\ref{sec:calc-resp-funct} for calculation of the time-domain response function.

    Figure~\ref{fig:response}(a) shows the frequency-domain form of the hopping response function
    $\hopresp^{(H)}(\omega)$ for various values of $\rho_{ex}$.  To give the peaks width, a numerical broadening is added, equivalent to multiplying the time-domain function by a decaying exponential.
    One clearly sees vibronic sidebands.
    Moreover, we find that these sidebands appear to survive at large $N$, as discussed in the next section.
    Such sidebands can in principle arise either from vibrational excitations on the charged molecule or in the active sector, we have checked that both processes occur.

    While there is a non-vanishing matrix element for occupying vibrational sidebands via hopping, as in the previous section, their contribution to the overall hopping rate is suppressed by a Boltzmann factor.   
    Prominent vibrational modes in organic materials
    are typically around $\wv\simeq 0.1$--$0.2$eV, which is larger than $k_B T$ at room temperature. 
    As such for these modes the transition to the ground state once again dominates.
    This is illustrated in Fig.~\ref{fig:response}(b), which shows the temperature dependence of the  contribution of the lowest energy final state to the overall hopping rate. 
    We denote the lowest energy final state $LP_0$ to indicate the vibrational ground state of the lower polariton.
    We show this for  various values of $\lamp$. 
    Note that when $\lamp=0$ (and so matches the configuration of the $\da$  molecules), there is a low probability of vibrational excitation at all temperatures.  
    Note also that in a material where there would be prominent vibrational modes comparable to $k_B T$, vibrational sidebands could become important.

    \subsection{Comparing hopping and absorption}
    \label{sec:abs}
    
    The appearance of sidebands of the lower polariton contrasts with the known behavior of the optical absorption~\cite{Cwik16,Herrera2016a,zeb2017}, where it is found that in the $N \to \infty$ limit there are no vibronic sidebands to the lower polariton.  
    The optical absorption spectrum, $\absorb(\omega)$, can be defined as the Fourier transform of the response function:
    \begin{equation}
    \absorb(t) \equiv
    \sum_j \absorb^j e^{-i \varepsilon_j t} =
    \mel{0}{\hat a(t) \hat a^\dagger(0) }{0},
    \end{equation}
    thus there is a close analogy to the hopping response function.   Since absorption only involves the active sector, states here are labeled by a single index $j$.   
    
\begin{figure}[htpb]
      \centering
    \includegraphics[width=1\columnwidth]{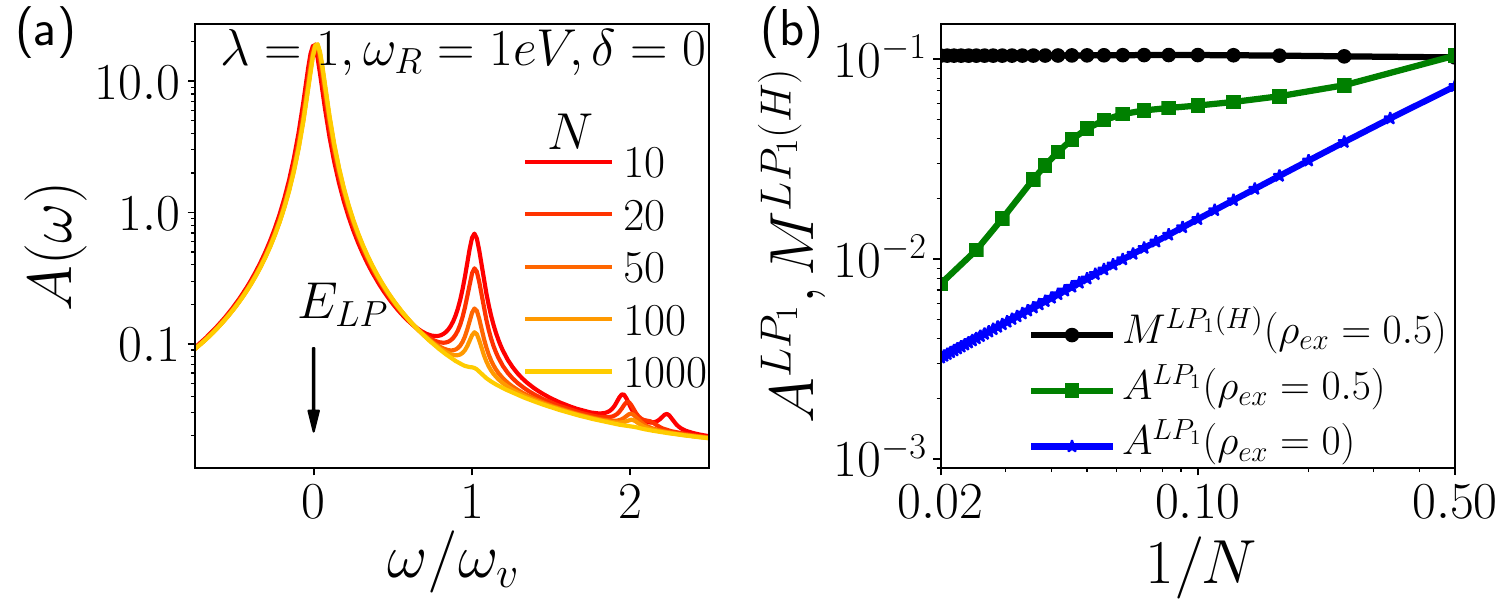}
    \caption{ 
      (a) Vibrational sidebands near the LP state for optical absorption. 
      Lines correspond to $N$ as indicated. Other parameters, including broadening, as in Fig.\ref{fig:response}.
      (b) Evolution of the weight of the first vibronic sideband---$(0-1)$ transition---for  optical absorption, $\absorb^{LP_1}$, and hopping, $\matelem^{LP_1(H)}$ vs $N$. Parameters as for (a), with $\rho_{ex}=0$ and $0.5$, respectively.
    }
      \label{fig:abs}
    \end{figure}

    Calculating the absorption spectrum using the permutation symmetric approach (see App.~\ref{sec:permutations}), one finds that
    vibronic sidebands of the lower polariton \emph{do} appear in the absorption spectrum when $N$ is small and $\wrr \gg \wv$, as shown in Fig.~\ref{fig:abs}(a). 
    That is, such states exist, but their weight $\absorb^j$ in the optical absorption vanishes as $1/N$ due to the delocalized nature of the polariton leading to a $1/N$ weight of the excitation on any single molecule.
    For hopping, the excitation process is localized to a single molecule.  This  allows the weight to survive.

    To verify the different dependence on $N$,  Fig.~\ref{fig:abs}(b),  compares the $N$ dependence of
    the probability to create a single vibrational excitation of the lower polariton, $LP_1$ in the two cases:
    $\matelem^{LP_1(H)}$ for hopping, and
    $\absorb^{LP_1}$,  for optical absorption. 
    This shows that the probability of creating a vibrational excitation survives at $N\to \infty$ for hopping, while it vanishes for optical absorption.   
    
    One may note that the hopping response and absorption response differ both in the operators acting on the states, $\hat V^c_{pq}$ vs $\hat a^\dagger$, and also in the initial state considered.  We defined absorption from the vacuum state, and hopping from a state with finite $N_{ex}$. 
    Figure~\ref{fig:abs}(b) also shows the result for optical absorption starting from a state with $\rho_{ex}=0.5$, and in this case the spectral weight of sidebands still vanishes at large $N$.   At larger $\rho_{ex}$ the sideband weight appears not to be suppressed over the range of $N$ accessible in our calculations.

    \section{Controlling hopping matrix elements with matter-light coupling}
    \label{sec:controlling}
    
    When including Boltzmann factors, the conclusion of the previous two sections is that at typical temperatures, the dominant hopping channel is the one which leaves the system unexcited---i.e.~$LP_0$ as the final state.
    Based on this, we focus the remainder of our discussion on the behavior of $R_{LP_0}$, and discuss how this rate is affected by matter-light coupling. 
    In particular, going beyond Refs.~\cite{HerreraPRL16,Schafer2019}, we focus on how the presence of a macroscopically occupied polariton mode changes the hopping rates. (For a related discussion in the context of vibrational strong coupling and vibrational polariton condensation, see Ref.~\cite{Pannir22}).
    We find that for sufficiently different $\lam$, $\lamp$, this change can be significant.
    The numerical results presented in this section are all derived using the methods of Appendix~\ref{sec:permutations}.
    
    \subsection{Evolution of hopping with matter-light coupling}

\begin{figure}[htpb]
      \centering
    \includegraphics[width=1\columnwidth]{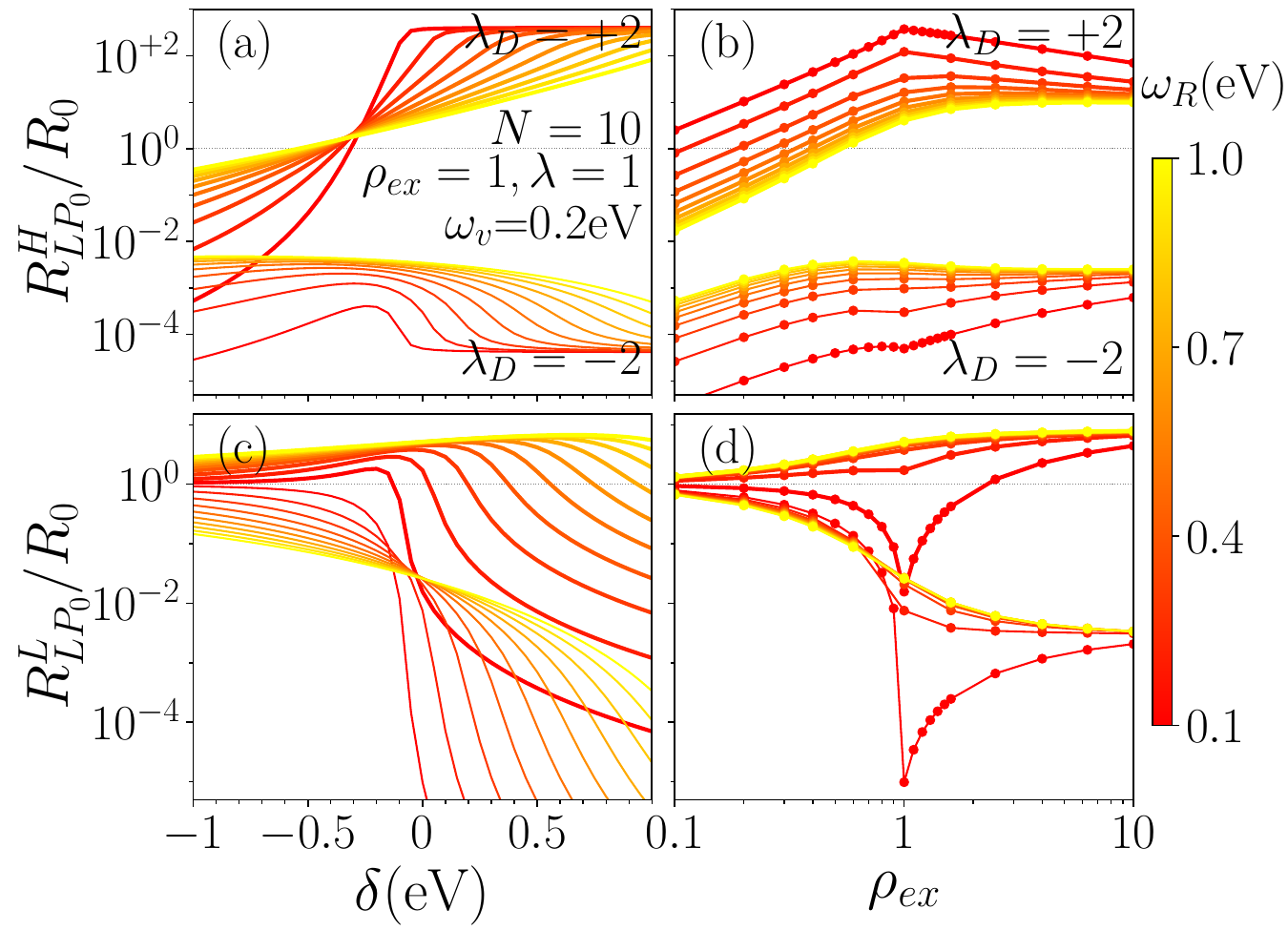} 
    \caption{Effects of matter-light coupling on normalized hopping rate
      $R^{(c)}_{LP_0}/R_0$ in the presence of the vibronic coupling.
      Left: (a,c)  vs cavity detuning $\delta\equiv \omega_c-\omega_0$ at $\rho_{ex}=1$.
      Right: (b,d) vs excitation density $\rho_{ex}$ at $\delta=0$.
      Top row (a,b) shows the HOMO channel, and bottom row (c,d) the LUMO channel.
      For all panels, various values of $\wrr$ are plotted, corresponding
      to the colorscale.  In addition, two sets of curves are shown for
      $\lamp=\pm2$ as labeled, with thicker (thinner) lines.
      We use $N=10$, all other parameters as in Fig.~\ref{fig:response}.}
      \label{fig:htc-rate}
    \end{figure}

    Figure~\ref{fig:htc-rate}
    shows the normalized channel-dependent hopping rates:
    \begin{displaymath}
    R^{(c)}_{LP_0}/R_0=\matelem^{LP_0(c)}/e^{-2\lamp^2},
    \end{displaymath}
    where the reference value $R_0$ is the hopping rate for zero matter-light coupling.  Since the hopping rate depends on the vibrational offset, the bare hopping rates differ for the HOMO and LUMO channels. We specifically chose $R_0$ to be the hopping in the LUMO channel.
    This ratio is shown as a function of cavity detuning $\delta\equiv \omega_c-\omega_0$ and excitation density at a range of matter-light couplings and at $\lamp=\pm 2$.

    The dependence on detuning, excitation density, and Rabi splitting in Fig.~\ref{fig:htc-rate} can be understood from considering two effects.  First is the variation of the fraction of excited molecules, $p_\ua$.  Hopping in the LUMO channel depends on $p_\da=1-p_\ua$, while hopping in the HOMO channel depends on $p_\ua$. 
    The second effect is the electronic-state-dependent vibrational offset,
    $\lam_\sigma$.
    Hopping in the LUMO channel depends on the difference $|\lamp-\lam_\da|$ while the HOMO channel depends on $|\lamp-\lam_\ua|$.   The larger this difference, the smaller the hopping rate.
    Both $p_\sigma$ and $\lambda_\sigma$ are affected by detuning, excitation density, and Rabi splitting~\cite{HerreraPRL16,zeb2017}.

    At large negative detuning excitations are mostly in the photon mode. Thus, all optically active molecules are in the $\da$ state.
    For these conditions, as seen in Fig.~\ref{fig:htc-rate}(a,c), hopping is only significant in the LUMO channel, and that channel recovers the rate in the absence of matter-light coupling.
    Increasing $\wrr$ transfers some excitations to the excited state, leading to enhancement of hopping in the HOMO channel,Fig.~\ref{fig:htc-rate}(a).  In the LUMO channel, Fig.~\ref{fig:htc-rate}(c),
    increasing $\wrr$ has opposite effects depending on the sign of $\lamp$. This dependence occurs because increasing $\wrr$ increases $\lam_\da$ (see discussion Sec.~\ref{sec:param-evol-with} below).  For $\lamp=+2$, increasing $\lam_\da$ enhances hopping, while for $\lamp=-2$, increasing $\lam_\da$ suppresses hopping.
    
    At positive detuning, excitations are favored in the molecules. Hopping is now significant in both channels.   In this case, increasing $\wrr$ decreases the fraction of excited molecules.  This effect suppresses hopping in the HOMO channel, and enhances it in the LUMO channel.
    One may however see that in the HOMO channel, Fig.~\ref{fig:htc-rate}(a), the behavior at large positive $\delta$ depends on the sign of $\lamp$.  In this case this occurs because increasing $\wrr$ \emph{decreases} $\lam_\ua$ (see Sec.~\ref{sec:param-evol-with}).
    
    Figure~\ref{fig:htc-rate}(b,d) shows the dependence of hopping on excitation density,  plotted at $\delta=0$.   Much of the behavior seen in this figure follows directly from the physics described above, with a general trend that increasing excitation density increases the fraction of excited active molecules.
    One may note that at small $\wrr$, the evolution of hopping is not monotonic with $\rho_{ex}$: there is a sharp minimum of $R^{(L)}_{LP_0}$ near $\rho_{ex}=1$, and a cusp in $R^{(H)}_{LP_0}$ at the same point.
    This effect follows directly from behavior of the probability of finding an active molecule in the excited state $p_\ua$.  The  probability $p_\ua$ first increases linearly with $\rho_{ex}$, reaches a maximum at $\rho_{ex}=1$, and then decreases toward $1/2$ at large $\rho_{ex}$.
    When $p_\ua=1$, the LUMO channel hopping contribution vanishes.
    The local maximum of $p_{\ua}$ at $\rho_{ex} \simeq 1$ has been observed and discussed previously~\cite{Eastham00,Eastham01}, as an effect which occurs at small $\wrr$ with positive detuning.
    Under such conditions, for $\rho_{ex}<1$ it is preferable to occupy the molecular states rather than the photon, so $p_\ua \simeq \rho_{ex}$. For $\rho_{ex}>1$ the photon must be occupied, and at very large $\rho_{ex}$, one then finds $p_{\ua}$ decreases to its asymptotic value of $1/2$, corresponding to the ground state in the presence of a large coherent photon field as was discussed in Sec.~\ref{sec:novib-large-rho}.
    In Fig.~\ref{fig:htc-rate}, while the bare detuning $\delta\simeq 0$, the vibronic reorganization energy reduces the exciton energy, so the effective detuning of the vibronically dressed transition is $\td = \delta + \lam^2 \wv > 0$.

    \subsection{Evolution of effective vibrational configuration}
    \label{sec:param-evol-with}
    
    \begin{figure}[htpb]
      \centering
    \includegraphics[width=1\columnwidth]{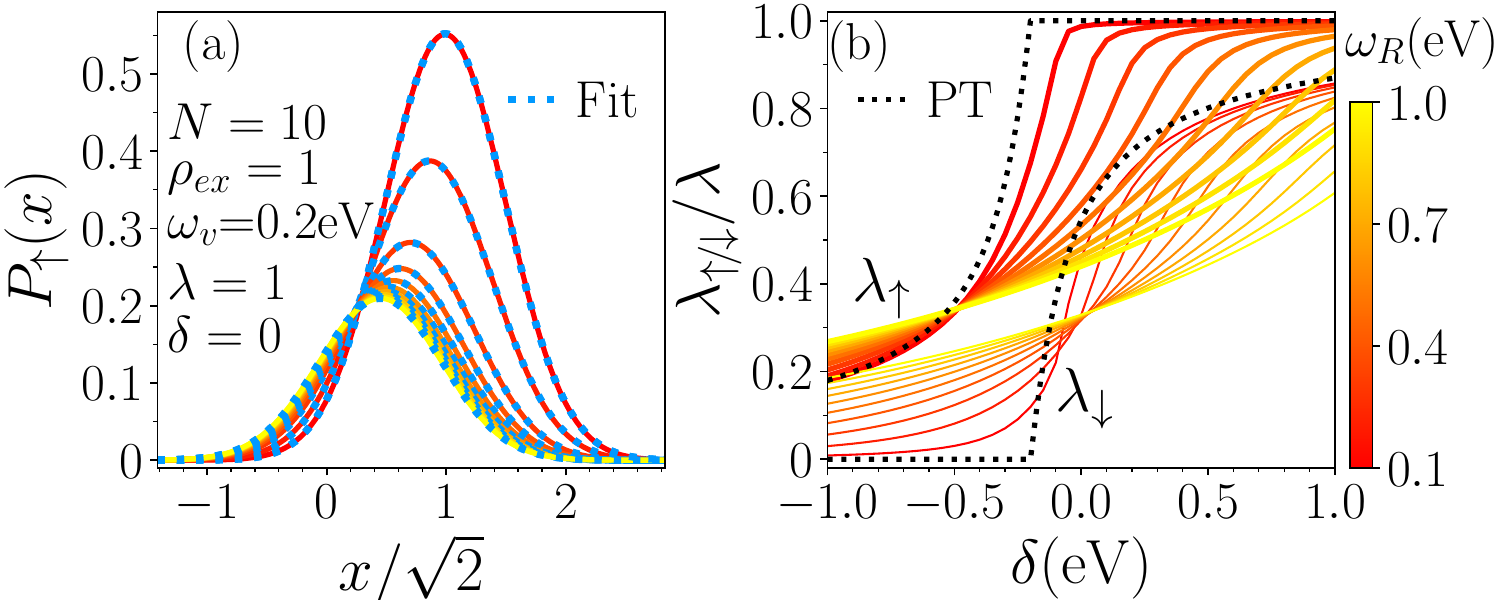}
    \caption{(a) Vibrational coordinate probability density $P_{\uparrow}(x)$ for a molecule being excited and having displacement $x$.  Red--yellow solid lines are for various values of $\wrr$ (see colorscale at right). Blue dotted lines are Gaussian fits.
      (b) Effective displacements, conditioned on ground or excited state of the given molecule, $\lam_{\da},\lam_{\ua}$.   Black dotted lines show the behavior in the limit $\wrr\to0$, found by  perturbation theory (Sec.~\ref{sec:pert-theory-at}).
      Parameters as indicated in panel (a).
    }
      \label{fig:fit}
    \end{figure}
    
    To further understand the behavior shown in Fig.~\ref{fig:htc-rate}(a), we discuss how the vibrational configuration of the lower polariton state evolves with coupling $\wrr$ and detuning $\delta$.
    The vibrational configuration for the singly excited state $N_{ex}=1$ was discussed extensively in Ref.~\cite{zeb2017}.  It was shown there that a Gaussian ansatz for the vibrational configuration was very good.  However,  results limited to $N_{ex}=1$ correspond to $\rho_{ex}\to 0$ at large $N$.  Here we extend the discussion to non-vanishing $\rho_{ex}$.
    
    Figure~\ref{fig:fit}(a) shows the probability density of the vibrational coordinate, $\hat{x}=(\hat b + \hat b^\dagger)/2$, in the polaritonic state with a particular electronic configuration $\sigma\in\{\ua,\da\}$, 
    \begin{displaymath}
    P_{\sigma}(x)=\sum_{\mu\nu} (\rho_{\sigma}^0)_{\mu\nu}
    \psi_\mu^\ast(x) \psi^{}_\nu(x),
    \end{displaymath}
    where $\psi_\mu(x)$ is the $\mu$th Gauss-Hermite function, and $\rho_\sigma^0$ is the reduced molecular density matrix element with electronic state $\sigma$.
    We clearly see that $P_{\sigma}(x)$ fits a Gaussian distribution very well.  Three fitting parameters are required: the overall weight which is $p_{\sigma}$, the effective width (trapping frequency, ${\wv}_{\sigma}$), and the vibrational displacement $\lambda_{\sigma}$.  We focus here on the displacements, $\lambda_{\sigma}$, as these have a strong effect on the transport.
    We present and discuss the other fitting parameters further in Appendix~\ref{sec:fitt-param-evol}, along with the dependence of all parameters on $\rho_{ex}$.

    The effective displacements $\lambda_\sigma$ extracted from the Gaussian fit are shown in Fig.~\ref{fig:fit}(b)
    as a function of $\delta$.  
    The behavior seen  can be explained as follows.  At large $\wrr$, the vibrational configuration is set by an average of the $\ua$ and $\da$ potential surfaces. As such, the results are similar for both displacements $\lambda_{\sigma}$, and evolve smoothly with $\delta$.
    This corresponds to the polaron decoupling limit~\cite{HerreraPRL16,wu2016,zeb2017}.
    
    At small $\wrr$ the results are more complicated, but, 
    as discussed in Sec.~\ref{sec:pert-theory-at}, can be calculated perturbatively in $\wrr$, as shown by the black dashed lines.
    For negative $\delta$ excitations are mostly in the photon, molecules in the $\da$ state, and so the displacement $\lambda_\da$ simply follows the configuration for unexcited molecules, so $\lambda_\da =0$.  In contrast, the behavior of $\lambda_\ua$  depends entirely on the weak excited molecule contribution to the ground state, and the vibronic configuration associated with that.    At large positive $\delta$, because we are considering $\rho_{ex}=1$, the scenario reverses. Now the ground state is purely excitonic, so $\lambda_\ua=\lam$,  and  $\lambda_\da$ depends on the state of the small fraction of unexcited molecules.  Note that the switch between the different regimes of detuning occurs when the effective detuning $\td$ discussed above crosses zero, i.e. at $\delta = -\lam^2 \wv$.
    
    Because the hopping rate depends exponentially on the difference $|\lam_{\sigma} - \lamp|$, the changes in $\lam_{\sigma}$ discussed here can be responsible for the order-of-magnitude changes in hopping rate seen in Fig.~\ref{fig:htc-rate}.

    \subsection{Perturbative calculation of displacements}
    \label{sec:pert-theory-at}
    
    As noted in the previous section, at small $\wrr$, one can calculate $\lam_\sigma$ perturbatively, corresponding to the dashed lines shown in Fig.~\ref{fig:fit}(b).  
    In this section,we provide details of this calculation.

    In the absence of matter-light coupling, the eigenstates
    of the HTC model are the vibrationally dressed versions of states with a fixed number $p$ of excited molecules and $N_{ex}-p$ photons.  We write this state as $\ket{(N_{ex}-p)_P; (p)_{ex}}$.
    Measuring energies with respect to the energy of the pure photon state $N_{ex} \omega_c$, these states have energies $E_{p,k}= - p\td + k\wv$ where the non-negative integer $k$ is the total number of vibrational quanta and  $\td \equiv \delta + \lam^2\wv$ as above.  
    In the following we will differentiate behavior depending on various conditions on $\td$ and $\rho_{ex}$. In each case we first discuss which state is the global minimum, and then consider the first-order change to that state due to matter-light coupling.

    \subsubsection{Negative detuning}
    
    For negative detuning, $\td<0$, at $\wrr=0$ the ground state is purely photonic.  Writing the vibrational state explicitly as $\ket{0_n}$ for the $n$th molecule, we have the zeroth order ground state:
    \begin{equation}
    \ket{\Psi^{0(0)}} = \ket{(N_{ex})_{P};\Da}  \otimes \bigotimes_n \ket{0_n}.
    \label{eq:lp0dn}
    \end{equation}
    To first order in matter-light coupling, this state couples to the one-exciton states with $k\geq 0$ vibrational excitations (denoted $1_k$ in the following):
    \begin{multline}
    \ket{\Psi^{1_k(0)}} = 
    \frac{1}{\sqrt{N}}\sum_{n=1}^{N}
    \ket{(N_{ex}-1)_{P};\ua_{n},\Da_{\neq n}}
    \\
     \otimes \hat D_n(\lam) \ket{k_n} \otimes
     \bigotimes_{m\neq n}\ket{0_m},
     \label{eq:k0dn}
    \end{multline}
    where $\hat D_n(\lam)=e^{\lam(\bd_n - \bn_n)}$ is the displacement operator for the $n$th molecule. Any non-negative integer $k$ is allowed. 
    
    The coupling between these states due to the matter-light coupling is
    \begin{multline}
    \mel{%
    \Psi^{1_k(0)}}{
    \frac{\wrr}{\sqrt{N}} \sum_n ( \hat \sigma_n^+ \hat a + \hat \sigma_n^- \hat a^\dagger )
    }{\Psi^{0(0)}} 
    \\
    =\wrr\sqrt{N_{ex}} \mel{k}{\hat D(-\lam)}{0}.
     \label{eq:hrdn}    
    \end{multline}
    The factor $\sqrt{N_{ex}}$ here comes from the  matrix elements of the photon annihilation operator.
    The matrix element of the displacement operator can be found from the overlap between a number state and a coherent state:
    \begin{displaymath}
    \mel{k}{\hat D(-\lam)}{0}=\frac{(-\lam)^k}{\sqrt{k!}}
    e^{-\lam^2/2}.
    \end{displaymath}
    We can thus write the ground state to first order in $\wrr$:
    \begin{multline}
    \ket{\Psi^0}=
    \ket{(N_{ex})_{P};\Da} \otimes \bigotimes_n \ket{0_n}
    \\+
    \twr\sum_{n=1}^{N}
    \ket{(N_{ex}-1)_{P};\ua_{n},\Da_{\neq n}}
    \\\otimes
    \sum_{k=0}^{\infty}
    \alpha^\ua_k
    \hat{D}_n(\lambda) \ket{k_n} \otimes \bigotimes_{m\neq n}\ket{0_{m}},
     \label{eq:lp1dn}
    \end{multline}
    where we have defined the coefficients
    \begin{displaymath}
     \alpha^\ua_k\equiv\frac{\mel{k}{\hat D(-\lam)}{0}}{\td - k\wv}
     =  -\int dx \mel{k}{\hat D(-\lam)}{0}e^{(\td - k \wv)x}.
    \end{displaymath}
    The second (integral) expression will be useful in the calculations below.
    
    When we consider the molecular density matrix conditioned on being in state $\kup$, the vibrational state of that molecule is  $\sum_k \alpha_k\hat D(\lambda)\ket{k}$.  From this, we can identify the parameter $\lambda_\ua$ by evaluating the expectation of the displacement operator, $\hat x = (\hat b + \hat b^\dagger)/2$ for the excited state:
    \begin{align}
      \lam_\ua &= \frac{\sum_{k,k^\prime} \alpha^\ua_{k^\prime} \alpha^\ua_{k}
        \mel{{k}^\prime}{\hat D(-\lambda)\hat x \hat D(\lambda)}{k}}{\sum_k |\alpha^\ua_k|^2}
      \nonumber\\&= \lambda +
      \frac{\sum_{k=1} \sqrt{k} \alpha^\ua_k \alpha^\ua_{k-1}}{\sum_k|\alpha^\ua_k|^2}.
    \end{align}
    By using the integral form of $\alpha_k^\ua$, one can evaluate the sums over $k$ to find:
    \begin{equation}
      \label{eq:deflu}
      \lam_\ua = \lam\left[1 -
        \frac{F_0(-\td/\wv, \lam) - F_0(1-\td/\wv, \lam)}{F_1(-\td/\wv, \lam)}
      \right],
    \end{equation}
    where we have defined:
    \begin{align}
      \label{eq:deff0}
      F_0(a,b)
      &\equiv \int_0^\infty dx \exp\left(-a x + b e^{-x} \right),
      \\
      \label{eq:deff1}
      F_1(a,b)
      &\equiv \int_0^\infty dx x \exp\left(-a x + b e^{-x} \right).
    \end{align}
    Closed (but complicated) forms for these integrals exist in terms of incomplete gamma functions and hypergeometric functions respectively.

    For $\lambda_\da$ the calculation is simpler.  Here we need the reduced density matrix conditioned on being in $\kdn$. In this case the state is just the unperturbed wavefunction, so (up to linear order in $\wrr$) $\lam_\da = 0$.

    \subsubsection{Positive detuning}
    
    For positive detuning,
    the zeroth order lowest polariton state $\ket{\Psi^{0(0)}} $ will be a highly excited molecular state.  When there are more excitations than molecules, $\rho_{ex}\geq 1$, this will be the maximally excited state with any extra excitations going into the photon mode. When there are fewer excitations than molecules, $\rho_{ex} < 1$, there are only $N_{ex} < N$ excited molecules, and no photons.  We consider these two cases separately.
    
    \paragraph{More excitations than molecules.}
    
    For this case, the zeroth order state is
    \begin{equation}
    \ket{\Psi^{0(0)}} = \ket{(N_{ex}-N)_{P};\Ua} \otimes
    \bigotimes_n D_n(\lambda) \ket{0_n}.
    \label{eq:lp0dp}
    \end{equation}
    The states this can couple to are the vibrational sidebands of states with  $N-1$ excitons and $N_{ex}-N+1$ photons, which we denote as:
    \begin{multline}
    \ket{\Psi^{1_k(0)}} = 
    \frac{1}{\sqrt{N}}\sum_{n=1}^{N}
    \ket{(N_{ex}-N+1)_{P};\da_{n},\Ua_{\neq n}}
    \\
     \otimes \ket{k_n}
     \otimes
     \bigotimes_{m\neq n} D_j(\lambda) \ket{0_m}.
     \label{eq:k0dp}
    \end{multline}
    Following the same procedure as for $\td<0$,
    we obtain the conditional state for an unexcited molecule
    is $\sum_k \alpha^\da_k \ket{k}$, where now we have
    \begin{displaymath}
      \alpha^\da_k \equiv
      \frac{\mel{k}{\hat D(\lam)}{0}}{\td + k\wv}.
    \end{displaymath}
    Using the same methods as above, this gives the effective displacement
    \begin{equation}
      \label{eq:defld}
      \lam_\da = \lam\left[
        \frac{F_0(\td/\wv, \lam) - F_0(1+\td/\wv, \lam)}{F_1(\td/\wv, \lam)}
      \right],
    \end{equation}
    with the same definitions in Eqs.~\eqref{eq:deff0},\eqref{eq:deff1}.

    For $\lambda_\ua$ we require the excited part of the state, which  is unaffected by the perturbation 
    so we have $\lam_\ua=\lam$.

    \paragraph{Fewer excitations than molecules.}
    \label{sec:ptexlt1}
    
    In this case, the maximum number of excited molecules is restricted to
    $N_{ex}<N$.  The zeroth order lowest polariton state  thus has $N_{ex}$ excitations in the excitons;
    $\ket{(0)_P; (N_{ex})_{ex}}$.  This expression introduces unexcited
    molecules in the zeroth order lowest polariton state.  This changes the expressions for the reduced density matrices, as both the $\ua$ and $\da$ states have a dominant contribution from the unperturbed wavefunction, i.e.
    $\lam_\da=0, \lam_\ua=\lam$.
    This case is not seen in Fig.~\ref{fig:fit}, since that figure shows $\rho_{ex}=1$.  Numerical results with $\rho_{ex}<1$ are shown in the Appendix~\ref{sec:fitt-param-evol}, in the top two rows of Fig.~\ref{fig:effpar}; these figures confirm the expected step-like behavior vs $\delta$.

    \section{Conclusions}
    We have found how a polariton condensate affects charge transport in organic materials, where transport proceeds by incoherent hopping.  
    To do this, we considered an extension of the Holstein--Tavis--Cummings model, incorporating charged states of molecules.
    This model provides a framework to understand incoherent charge transport in systems with strong matter-light coupling.
    We have presented exact numerical results, based on the use of permutation symmetry~\cite{zeb2017,Zeb2022:FockMap}, which scales polynomially with the number of molecules $N$.
    We have shown that in several limiting cases, these results can also be understood by analytic expressions that hold at all $N$.
    By combining these results, we demonstrate that the permutation symmetric approach is capable of showing behavior consistent with the large $N$ asymptotic limit.

    When a charge hops between molecules, various excited states can be created, by transferring a lower polariton to an upper polariton or dark state, or creating vibrational sidebands.  
    While these processes can have significant matrix elements, the ground state process dominates the hopping at relevant temperatures. Even when remaining in the ground state, the hopping rate depends strongly on the condensate density, detuning, and matter-light coupling, through modification of the effective vibrational configuration of those molecules forming the polariton condensate.
    This changes the overlap between the vibrational configurations of the molecules between which the charge hops, leading to dramatic changes of the hopping rates.  
    
    One question for future work is to explore models beyond that considering a single vibrational mode, to consider the role of low frequency vibrational and rotational modes. 
    Another  possible future direction would be to explore the consequences of our results for producing an electrically pumped polariton condensate~\cite{schneider2013,Bhattacharya2013} in an organic microcavity. 
    Understanding and exploiting the strong dependence of transport on matter-light coupling and excitation density may be significant for such experiments.
    
    \begin{acknowledgments}
      The authors acknowledge financial support from EPSRC program ``Hybrid Polaritonics'' (EP/M025330/1) and an ESQ fellowship of the Austrian Academy of Sciences (\"OAW) (PK).
      MAZ thanks Rukhshanda Naheed for fruitful discussions.
    \end{acknowledgments}
    
    \appendix

    \section{Permutation symmetric bases for exact diagonalization}
    \label{sec:permutations}
    
    In this appendix we describe the numerical method used  to calculate behavior at finite $\rho_{ex}$. 
    This is based on exploiting permutation symmetry of the Holstein--Tavis--Cummings  model under interchange of molecules.
    This permutation symmetry, in the single excitation subspace, was described in Ref.~\cite{zeb2017,Zeb2022:FockMap} (see in particular the Supplementary Information of that reference), and in Ref.~\cite{Zeb2022:FockMap}.  
    Here we describe how to extend these ideas to the case with multiple excited molecules.  To make this appendix self contained, we include here some points discussed in those previous works.
    
    The main point to note is that, in general, there are many states that are equivalent when transformed by interchanging molecules. 
    Our approach is based on keeping  a single representative state for all states related to it by such permutations.  
    We will first discuss how we label these representative states in Sec.~\ref{sec:basis-set}, we then discuss how to write the Hamiltonian in terms of these basis states in Sec.~\ref{sec:htc-hamilt-perm}. Section~\ref{sec:cond-reduc-vibr} shows how to extract information about the vibrational state of a given molecule, while Sec.~\ref{sec:calc-resp-funct} discusses calculation of the response functions shown in Sec.~\ref{sec:hop-vib}.
    
    \subsection{Permutation symmetric basis set}
    \label{sec:basis-set}
    
    In this section we define the permutation symmetric basis set.
    We first consider the electronic and photonic states alone, temporarily ignoring vibrations.  In such a case, we know that the Tavis--Cummings model could be efficiently solved using collective spin operators.  However, to provide the framework for the general case, it is useful to consider this explicitly through permutations.
    
    For $N$ molecules and $N_{ex}$ excitations, the number of excited molecules can range between zero and $\text{min}(N_{ex},N)$.
    If there are $p$ molecules excited, there are $N-p$ molecules unexcited,
    and $N_{ex}-p$ photons;  we can write the excitonic part of this state in
    the form:
    \begin{multline}
      \ket{(p)_{ex}}
      \equiv
     \frac{1}{\sqrt{{}^{N}C_{p}}} \times\\
      \sum_{n_1>n_2>\ldots >n_p}^N\!\!\!\!\!\!\!\!
      \ket{\ua_{n_1}\ua_{n_2}\ldots \ua_{n_p}}\ket{\Da_{\neq n_1,n_2,\ldots ,n_p}},
    \end{multline}
    where $\Da_{\neq n_1,n_2,\ldots ,n_p}$ denotes the state of the unexcited molecules.
    
    We next include  vibrations.  
    We first consider the vibrational state of the excited molecules. 
    The set of unexcited molecules can then be treated in a similar fashion.
    Given $p$ excited molecules, there exist a set of vibrational states which
    are related by permuting the vibrational quantum numbers on each molecule.
    If we denote $\{\nu\}$ as the set of vibrational quantum numbers---i.e. the set of numbers of excitations, then the
    permutation symmetric superposition of such states
    $\ket{\symset_{p}\{\nu\}}$,
    is given by,
    \begin{equation}
    \label{eq:perm_state}
    \ket{\symset_{p}\{\nu\}} \equiv 
    \frac{1}{{\sqrt{\cntset_p(\{\nu\})}}}
    \sum_{\text{P}} \ket{\text{P}[\{\nu\}]}
    ,
    \end{equation}
    where $P$ indicates a permutation,
    and
    $\cntset_p[\{\nu\}]$ counts the number of distinct permutations which will depend on the pattern of occupations in $\{\nu\}$. 
    If we label the frequency $f_{\nu_n}$ as the number of times each value $\nu_n$ appears in the set $\{\nu\}$, then the number of permutations is the multinomial coefficient  $\cntset_p(\{\nu\})=p!/(\prod_n f_{\nu_n}!)$. 
    For example, 
    for the set of occupations $\{0112\}$, the frequencies are $1,2,1$ and so
    $\cntset_4(\{0112\})=12$,
    and the permutation symmetric state is:
    \begin{displaymath}
    \ket{\symset_{4}\{0112\}} \equiv 
    \frac{
    \begin{aligned}
    \bigl(&\ket{0112} + \ket{1012} + \ket{1102} + \ket{1120} \\ 
    +& \ket{0211}  + \ket{2011}  + \ket{2101}  + \ket{2110} \\ 
    +& \ket{0121}+ \ket{1021}+ \ket{1201}+ \ket{1210} \bigr)
    \end{aligned}
    }{\sqrt{12}}.
    \end{displaymath}

    We can write the  permutation symmetric state
    for the unexcited molecules, with vibrational configuration $\{\mu\}$,
    in the same fashion,
    $\ket{\symset_{N-p}\{\mu\}}$.
    Putting together the photon, electronic, and vibrational states, we can
    denote a general state in the following form:
    \begin{multline}
      \label{eq:def-state}
    \ket{\{\nu\}_p\{\mu\}_{N-p}} \equiv\\
    \ket{(N_{ex}-p)_P;(p)_{ex}}
    \otimes
    \ket{\symset_{p}\{\nu\}}
    \otimes
    \ket{\symset_{N-p}\{\mu\}}.
    \end{multline}
    Here, the first ket labels the photon and electronic states, while the second and third are the vibrational states of the excited and unexcited molecules which have configurations $\{\nu\}$ and $\{\mu\}$ respectively.
    In the following it is necessary to define a canonical representative configuration of $\{\nu\},\{\mu\}$, so we can ensure to count each equivalent configuration only once.
    We choose our canonical representation so that the occupations are in increasing order, such as in the example $\{0112\}$ written above.

    To perform numerical calculations, the vibrational number states need to be
    truncated.  We thus introduce the vibrational cutoff $\maxnu$, such that
    $\nu_{n}, \mu_{n} \in [0,\maxnu]$.  In the figures shown, we always take $M$ greater than $5$, and in all cases we checked the results were converged with the value of $\maxnu$ used.

    The size of the permutation symmetric 
    subspace is exponentially smaller than the
    full Hilbert space.
    The total number of distinct permutation symmetric vibrational states
    for $p$ excited molecules
    is ${}^{\maxnu+p}C_{\maxnu}$ compared to a total of $(\maxnu+1)^p$ states.
    This counting comes from the number of ways to pick $p$ numbers in the range $[0,\maxnu]$ ignoring order.
    The size of the permutation symmetric 
    space is therefore
    $\sum_{p=0}^{\min(N_{ex},N)} \left[ {}^{\maxnu+p}C_{\maxnu} \times {}^{\maxnu+N-p}C_{\maxnu} \right]$
    which increases only polynomially with $N$, much slower than the exponential size of the full Hilbert space $2^N\times(\maxnu+1)^N$.
    This far better scaling makes it possible 
    to calculate the lowest polariton eigenstate of Holstein--Tavis--Cummings model for  values of $N, N_{ex}, \maxnu$ that are large enough to identify the behavior in the thermodynamic limit.
    The downside of this approach is that, as discussed next, 
    the calculation of the matrix elements of the Hamiltonian 
    and the reduced vibrational density matrices are 
    not trivial.

    \subsection{HTC Hamiltonian in the permutation symmetric basis set}
    \label{sec:htc-hamilt-perm}
    
    In this section we discuss how to write the HTC Hamiltonian in the permutation symmetric state space, considering each term in turn.
    
    \subsubsection{Diagonal terms}
    
    The diagonal terms of Holstein--Tavis--Cummings model are straightforward.
    In the state $\ket{\{\nu\}_p\{\mu\}_{N-p}} $, the operators
    $\ad \an$ and $ \sum_n \hat \sigma^+_n \hat \sigma^-_n$
    count number of cavity photons ($N_{ex}-p$) and excited molecules ($p$), respectively.
    The vibrational excitation number, $\sum_n \hat b^\dagger_n \hat b^{}_n$,
    becomes $\sum_{n=1}^{p} \nu_n+\sum_{m=1}^{N-p}\mu_m$.

    \subsubsection{Vibrational coupling}
    
    The term coupling the electronic and vibrational states,
    \begin{math} 
    \sum_n\hat\sigma^+_n \hat \sigma^-_n ( \hat b^\dagger_n + \hat b^{}_n ), \nonumber
    \end{math}
    acts only on the excited molecules
    and involves matrix elements of the position operator.
    As such, we must find the off-diagonal matrix elements in the $\ket{\{\nu\}_{p}}$ subspace.  
    
    We consider the vibrational creation operator term,  which we can write as
    $\sum_n \hat\sigma^+_n \hat \sigma^-_n \hat b^\dagger_n=\sum_{n\in \text{excited}}\hat b^\dagger_n$; the annihilation term follows by conjugation. 
    By choosing the representative state to have molecules $n=1 \ldots p$ excited, the matrix element can be written explicitly as a sum over permutations:
    \begin{multline} 
    \label{eq:vibmel}
    \bra{\{\nu^\prime\}_p\{\mu\}_{N-p}}
      \sum_{n=1}^p\hat b^\dagger_n
     \ket{\{\nu \}_p\{\mu\}_{N-p}} 
     =\\
     \sum_{P,P^\prime} 
     \frac{\bra{P^\prime[\{\nu^\prime_1\nu^\prime_2\ldots \nu^\prime_p\}]}}{\sqrt{\cntset_p(\{\nu^\prime\})}}
     \sum_{n=1}^p \hat b^\dagger_n
     \frac{\ket{P[\{\nu_1\nu_2\ldots \nu_p\}]}}{\sqrt{\cntset_p(\{\nu\})}}.
     \end{multline}
     
    Let us consider a single term
    $ \hat b^\dagger_n
     \sum_{P} \ket{P[\{\nu_1\nu_2\ldots \nu_p\}]}$.
    For a given permutation $P$, if we write $\nu_{P(n)}$ for the vibrational quantum number of the $n$th molecule after that permutation, then this term will give an expression of the form
    $\sqrt{\nu_{P(n)}+1} $ times the state with $\nu_{P(n)}\to \nu_{P(n)} +1$. 
    For this to have a non-zero overlap with 
    $\bra{P^\prime[\{\nu^\prime_1\nu^\prime_2\ldots \nu^\prime_p\}]}$
    for at least one
    permutation $P^\prime$ we require that
    $\{\nu^\prime\}$ 
    is the same as $\{\nu\}$ 
    except $\nu_{P(n)}\to\nu_{P(n)} + 1$;
    i.e., the \emph{multiset differences} are
    $\{\nu \}_p \setminus \{\nu^\prime \}_p = \{\nu_{P(n)}\}$
    and 
    $\{\nu^\prime \}_p \setminus \{\nu \}_p=\{\nu_{P(n)}+1\}$.
    
    Since Eq.~\eqref{eq:vibmel} involves the sum over all active molecules $n$, we may write expressions in a way independent of molecule labels.  
    The matrix element in
    Eq.~\eqref{eq:vibmel} will be non-zero if and only if there exists $\nu_{\ast}$ such that
    the multiset differences are
    $\{\nu \}_p \setminus \{\nu^\prime \}_p = \{\nu_\ast\}$
    and 
    $\{\nu^\prime \}_p \setminus \{\nu \}_p=\{\nu_\ast+1\}$.
    If so, every ket in the permutation $P$
    finds its dual in $P^\prime$.
    In other words, the only difference between these two configurations is
    that their frequencies of $\nu_\ast$ and $\nu_\ast+1$ are different and related by
    $f_{\nu_\ast}(\{\nu\})=f_{\nu_\ast}(\{\nu^\prime\}) + 1 $
    and $f_{\nu_\ast+1}(\{\nu\})=f_{\nu_\ast+1}(\{\nu^\prime\}) - 1$.
    Since, there are
    $\cntset_p(\{\nu\}) $ permutations of
    $\{\nu\}$,
     we will get
    $ \cntset_p(\{\nu\})\times \sqrt{\nu_\ast+1}$ for one such term.
    Noting that the element $\nu_\ast$ may occur multiple times in the set $\{\nu\}$, and that its frequency is $f_{\nu_\ast}(\{\nu\})$,
    the matrix element then becomes
    \begin{align} 
    \lefteqn{
    \bra{\{\nu^\prime\}_p\{\mu\}_{N-p}}
      \sum_{n=1}^p\hat b^\dagger_n
      \ket{\{\nu \}_p\{\mu\}_{N-p}}}
    \nonumber\\&=
    \sqrt{
    \frac{\cntset_p(\{\nu\})}
    {\cntset_p(\{\nu^\prime\})}
    }
    f_{\nu_\ast}(\{\nu\})
    \sqrt{\nu_\ast+1}
    \nonumber\\&=
    \sqrt{(\nu_\ast+1)f_{\nu_\ast}(\{\nu\})(f_{\nu_\ast+1}(\{\nu\})+1)},
    \end{align}
    where the last expression  uses the  definition of ${\cntset_p(\{\nu\})}$  in Eq.~\eqref{eq:perm_state}.

    \subsubsection{Matter-light coupling}
    
    The matter-light coupling, 
    \begin{math}
      \sum_n\left(\hat \sigma^+_n \hat a^{} + \hat \sigma^-_n \hat
        a^\dagger\right)
    \end{math},
    couples states with $p$ excited molecules
    to those with $p\pm1$ excited molecules.
    While this term does not change the vibrational state, the labeling of vibrational states before and after differs, due to the changing excitation number.
    We focus on the photon creation term,
    $\sum_n \hat \sigma^-_n \hat a^\dagger$, 
    the other term follows by conjugation.
    
    We first write out the matrix element in terms of the explicit states, Eq.~(\ref{eq:def-state}),
    \begin{multline} 
    \lefteqn{
    \bra{\{\nu^\prime\}_{p-1}\{\mu^\prime\}_{N-p+1}}
    \sum_n \hat \sigma^-_n \hat a^\dagger
     \ket{\{\nu\}_{p}\{\mu\}_{N-p}} }
    \\=
    \mel{(N_{ex}-p+1)_P  ;{(p-1)_{ex}}
    }{%
      \sum_n \hat \sigma^-_n \hat a^\dagger
    }{(N_{ex}-p)_P ;{(p)_{ex}}
    }
    \\
    \times
    \braket{
    {\symset_{p-1}\{\nu^\prime\}}
    ;{\symset_{N-p+1}\{\mu^\prime\}}
    }{{\symset_{p}\{\nu\}}
    ;{\symset_{N-p}\{\mu\}}
    }
    \\
    =
    \sqrt{(N_{ex}-p+1)(p)(N-p+1)}  \times\mathcal{O}_V.
      \end{multline}
    Here,
    $\mathcal{O}_V\equiv 
    \braket{
    {\symset_{p-1}\{\nu^\prime\}}
    ;{\symset_{N-p+1}\{\mu^\prime\}}
    }{{\symset_{p}\{\nu\}}
    ;{\symset_{N-p}\{\mu\}}
    }
    $ is the vibrational overlap.
    It will be non-zero
    only when
    the vibrational states
    of all molecules
    in the ket are the same as those in the bra.
    This means that by taking a single element $\nu_\ast$ out
    of $\{\nu\}_p$, the rest should become equal to $\{\nu^\prime\}_{p-1}$,
    and, similarly, 
     by adding the same element $\nu_\ast$ to $\{\mu\}_{N-p}$
     should make $\{\mu^\prime\}_{N-p+1}$.
    In such a case, the overlap is given by counting the number of non-zero overlapping elements, and scaling by the normalization of the initial and final states:
    \begin{align} 
    \mathcal{O}_V
    &= 
    \frac{\cntset_{p-1}(\{\nu^\prime\})}{\sqrt{
    \cntset_{p-1}(\{\nu^\prime\}) 
    \cntset_{p}(\{\nu\}) 
    }}
    \frac{\cntset_{N-p}(\{\mu\})}{\sqrt{
    \cntset_{N-p+1}(\{\mu^\prime\}) 
    \cntset_{N-p}(\{\mu\})
    }}
    \nonumber\\&= 
    \sqrt{
         \frac{f_{\nu_\ast}(\{\nu\})}{p}~
         \frac{f_{\nu_\ast}(\{\mu^\prime\})}{N-p+1}}.
    \end{align}
    Because the initial and final states here involve different numbers of excited molecules, we need to establish a map between the indexing of states in the two manifolds.  We will denote this map $\maptop$.
    We first introduce $\indset_p(\{ \nu\})$ as the index 
    of the configuration  $\{\nu\}$ in the manifold with $p$ excitations (see below).  
    We can then define a map $\maptop$ from the pair of
    integers $(\nu_\ast, \indset_{p-1}(\{ \nu^\prime \}))$ which identifies which state
    $\indset_p(\{\nu\})$ one achieves when adding $\nu_\ast$ to the set $\{ \nu^\prime \}$. A similar map, $\maptonp$, results from the second condition. 
    
    \subsubsection{Index and mapping}
    
    We choose to index the configurations $\{\nu\}$ in lexicographic order, starting from $\indset_p(\{0,0,\ldots,0\})=0$.  An explicit expression for $\indset_p(\{ \nu\})$ can then be found as follows.     Recall first that our representative patterns $\{\nu_1,\nu_2,\nu_3,\ldots,\nu_p\}$ are arranged in increasing order, $\nu_1 \leq \nu_2 \leq \nu_3 \ldots$.  To find the index $\indset_p(\{ \nu\})$ we must count the patterns that occur \emph{before} the current pattern.  This can be done recursively, by considering each successive label $\nu_n$, starting from $n=1$.  That is, the number of patterns preceding $\{\nu_1,\nu_2,\nu_3,\ldots,\nu_p\}$ is given by the sum of the following: the number of patterns preceding $\{\nu_1,\nu_1,\nu_1,\ldots,\nu_1\}$, the number of patterns between $\{\nu_1,\nu_1,\nu_1,\ldots,\nu_1\}$ and $\{\nu_1,\nu_2,\nu_2,\ldots,\nu_2\}$, the number of patterns between $\{\nu_1,\nu_2,\nu_2,\ldots,\nu_2\}$ and $\{\nu_1,\nu_2,\nu_3,\ldots,\nu_3\}$, etc.  Each of these expressions follows the same general form, as the $n$th such term corresponds to enumerating the allowed ``previous'' values of $\nu_n$, i.e. the set $\nu^\prime$ satisfying $\nu_{n-1} \leq \nu^\prime < \nu_n$, and then
    counting the number of ways of assigning a limited set of indices $[\nu^\prime, \maxnu]$ to the remaining $p-n$ sites.  This counting is given by the same combinatoric factor as occurs when counting the total set of patterns.  We thus have:
    \begin{eqnarray}
    \indset_p(\{ \nu\})
    =\sum_{n=1}^{p} \left[
    \sum_{\nu^\prime=\nu_{n-1}}^{\nu_n-1}
    {}^{(\maxnu-\nu^\prime)+(p-n)}C_{p-n}
    \right].
    \end{eqnarray}
    We note that for $n=1$, the lower limit of the sum over $\nu^\prime$ should be taken as $\nu_0\equiv 0$, since there is no previous site to constrain the lower limit of $\nu^\prime$.  We note also that if $\nu_{n-1}=\nu_n$ there are no terms in the inner sum so it gives zero.
    
    With such an explict expression for the index, the construction of the map $\maptop$ becomes straightforward.   One first enumerates (in lexicographic order) the patterns $\{\nu^\prime\}=\{\nu_1, \nu_2, \nu_3, \ldots \nu_{p-1}\}$.  For each such pattern one then enumerates over the ``extra'' label $\nu_\ast \in [0, \maxnu]$,  and constructs and sorts the set
    $\{\nu\}=\{\nu^\prime\} \cup \{\nu_\ast\}$.
    One then finds the index of this new pattern,  $\indset_p(\{ \nu\})$, providing the map.
    Examples of this map, along with an alternate method of its construction by identifying a recursive pattern, can be founds in Ref.~\cite{Zeb2022:FockMap} and the associated code~\cite{FockMapCode}. 
    
    \subsection{Reduced vibrational density matrices} 
    \label{sec:cond-reduc-vibr}
    In this section, we discuss how one can determine the reduced vibrational density matrices using the permutation symmetric space.  These density matrices can be used to calculate hopping rates to the vibrational ground state.
    In section~\ref{sec:calc-resp-funct} below, we discuss how to calculate the hopping rates in the general case.
    
    We can write eigenstate $r$  as follows:
    \begin{equation}
      |\Psi^r\rangle = 
      \sum_{p=1}^{\text{min}(N_{ex},N)}\!\!\!\!\!\!\!\!
      \sum_{\substack{k_p\equiv\indset_p{(\{\nu\})}\\l_{N-p}\equiv\indset_{N-p}{(\{\mu\})}}}\!\!\!\!
      \psi^r_{k_p,l_{N-p}}  \ket{\{\nu\}_p\{\mu\}_{N-p}},
    \end{equation}
    where $k_p,l_{N-p}$ index the vibrational patterns of the excited and unexcited molecules, as introduced above. A crucial step to calculating observables is to define an object which we will denote as $\rho_{\sigma}^r$.
    This object, which in general is \emph{not} a density matrix, describes the vibrational configuration of a single molecule associated with coherence between the ground state $\ket{\Psi^0}$ and the state $\ket{\Psi^r}$, conditioned on the molecule in question being in the $\sigma\in{\ua,\da}$ state.  This is defined by taking a trace over the electronic and vibrational configurations of all molecules other than the one in question.  This can be written as:
    \begin{equation}
      (\rho^{r}_{\sigma})_{\nu,\nu^\prime}=
      \braket{\sigma \nu }{\Psi^0}
      \braket{\Psi^r}{\sigma  \nu^\prime}
    \end{equation}
    Here we suppressed molecule labels (since states are permutation symmetric), and $\nu, \nu^\prime$ denote vibrational quantum numbers of the molecule in question.
    As noted above, unless $r=0$, this object is not a reduced density matrix.

    To evaluate this, we need to trace out the vibrational state of the $N-1$ other
    molecules. This can be done using the maps    $\maptop$ as defined above or $\maptonp$,
    applied respectively to the $p$ excited molecules or to the $N-p$ unexcited molecules.  We discuss these two cases in turn.

    \subsubsection{Excited molecules, \texorpdfstring{$\rho_{\ua}^r$}{}}

    To find the element $({\rho_{\ua}^r})_{\nu,\nu^\prime}$, 
    we need to find all pairs
    of states with $p$ excited molecules 
    which are reduced to the same $p-1$
    molecule state when $\nu,\nu^\prime$ are taken out. 
    For example, if we denote
    $k^{}_{p} = \indset_{p}(\{\nu\})$
    and 
    $k^{\prime}_{p} = \indset_{p}(\{\nu^\prime\})$
     as the indices of a pair of states $\{\nu\}$ and $\{\nu^\prime\}$ of $p$ excited molecules,
     that reduce to the same state $\{\nu^{\prime\prime}\}$ of $p-1$ excited molecules with index
    $j^{}_{p-1} = \indset_{p-1}(\{\nu^{\prime\prime}\})$,
    we can write
    \begin{equation}
      \label{eq:define_p_map}
      \begin{gathered}
      k^{}_p    = \maptop(\nu,      j_{p-1}), \\
      k^\prime_p = \maptop(\nu^\prime,j_{p-1}).
      \end{gathered}
    \end{equation}
    With these maps,  we can then trace over $j_{p-1}$, describing the state of
    the other excited molecules.
    The trace over the set of unexcited molecules is trivial.

    \begin{widetext}
      Taking $k^{}_p, k^\prime_p$ as defined by Eq.~\eqref{eq:define_p_map}  we find $(\rho_{\ua}^r)_{\nu,\nu^\prime}$ takes the form:
    \begin{align}
      \label{eq:cdms}
      (\rho_{\ua}^r)_{\nu,\nu^\prime} &=
      \sum_{p=1}^{\min(N_{ex},N)} 
      \frac{{}^{N-1}C_{p-1}}{\sqrt{{}^{N}C_{p}~ {}^{N}C_{p}}}
        \sum_{l_{N-p}=1}^{\mathcal{N}_{N-p}}
        ~
      \sum_{j_{p-1}=1}^{\mathcal{N}_{p-1}}
      \frac{
      \psi^{0 \ast}_{k_p,l_{N-p}} \psi^{r}_{k^\prime_p,l_{N-p}}
       \cntset_{p-1}(j_{p-1})
      }{\sqrt{\cntset_{p}(k_{p}) \cntset_{p}(k^\prime_{p})}}
      \nonumber\\
      &=
        \sum_{p=1}^{\min(N_{ex},N)} 
       \frac{p}{N} 
        \sum_{l_{N-p}=1}^{\mathcal{N}_{N-p}}
        ~
      \sum_{j_{p-1}=1}^{\mathcal{N}_{p-1}}
      \frac{
      \psi^{0 \ast}_{k_p,l_{N-p}} \psi^{r}_{k^\prime_p,l_{N-p}}
       \cntset_{p-1}(j_{p-1})
      }{\sqrt{\cntset_{p}(k_{p}) \cntset_{p}(k^\prime_{p})}}.
    \end{align}
    Here $\mathcal{N}_{q}$ is the total number of the permutational
    symmetric vibrational basis states
    involving $q$ molecules. 
    \end{widetext}
    
    The factors in the denominator come from the normalization of the
    permutation symmetric basis states. 
    The factor ${}^{N-1}C_{p-1}$ in the
    numerator counts how many terms in the permutation symmetric superposition
    of excited molecules contain the specific molecule under consideration. The
    final factor $\cntset_{p-1}(j_{p-1})$ counts the number of matching terms
    in the permutation symmetric superposition of the vibrational states
    $k_{p},k^\prime_{p}$---and thus give unit overlap---after taking out the
    vibrational states of our subject molecule.
    
    \subsubsection{Unexcited molecules, \texorpdfstring{$\rho_{\da}^r$}{}}
    We can use a similar approach to calculate  $\rho_{\da}^r$.
    The indices of the basis states
    with $N-p$ and $N-p-1$ unexcited molecules
    can be written as,
    \begin{equation}
      \label{eq:define_np_map}
      \begin{gathered}
       k^{}_{N-p}    = \maptonpm(\nu,      j_{N-p-1}), \\
      k^\prime_{N-p} = \maptonpm(\nu^\prime,j_{N-p-1}).
      \end{gathered}
    \end{equation}
      \begin{widetext}
    Taking $k^{}_{N-p}, k^\prime_{N-p}$ defined by Eq.~\eqref{eq:define_np_map} the matrix elements of
    ${\rho_{\da}^r}$ can then be written as,
    \begin{align}
      \label{eq:cdms2}
      (\rho_{\da}^r)_{\nu,\nu^\prime} &=
      \sum_{p=0}^{N-1}  
       \frac{{}^{N-1}C_{N-p-1}}{\sqrt{{}^{N}C_{p}~ {}^{N}C_{p}}}
        \sum_{l_{p}=1}^{\mathcal{N}_{p}}
        ~ 
      \sum_{j_{N-p-1}=1}^{\mathcal{N}_{N-p-1}}
      \frac{
      \psi^{0 \ast}_{l_{p},k_{N-p}} \psi^{r}_{l_{p},k^\prime_{N-p}}
       \cntset_{N-p-1}(j_{N-p-1})
      }{\sqrt{\cntset_{N-p}(k_{N-p}) \cntset_{N-p}(k^\prime_{N-p})}}
      \nonumber\\
      &=
      \sum_{p=0}^{N-1}  \frac{N-p}{N}  
        \sum_{l_{p}=1}^{\mathcal{N}_{p}}
        ~ 
      \sum_{j_{N-p-1}=1}^{\mathcal{N}_{N-p-1}}
      \frac{
      \psi^{0 \ast}_{l_{p},k_{N-p}} \psi^{r}_{l_{p},k^\prime_{N-p}}
       \cntset_{N-p-1}(j_{N-p-1})
      }{\sqrt{\cntset_{N-p}(k_{N-p}) \cntset_{N-p}(k^\prime_{N-p})}}.
    \end{align}
    
    \end{widetext}
    
    \subsection{Hopping response function}
    \label{sec:calc-resp-funct}

    In this section we discuss how to calculate the hopping response function,
    $\hopresp^{(c)}(\omega)$, defined in Eq.~\eqref{eq:hoppingresp}.  We describe two approaches below; the first is the one we use numerically.  The second shows how this quantity can in principle be related to the quantities introduced in the previous section.
    
    \subsubsection{Time evolution}
    We may find the hopping response function, $\hopresp^{(c)}(\omega)$,
    by computing $\hopresp^{(c)}(t)$ using direct time evolution.
    We start with an initial state, $\ket{\Psi_{\aset^\prime \cup \{q\}}^0\Phi_p^0}$, 
    where $\ket{\Psi_{\aset^\prime \cup \{q\}}^0}$ and $\ket{\Phi_p^0}$ are the ground states of Holstein--Tavis--Cummings model in the active sector, $\aset^\prime \cup \{q\}$, and the Holstein model on molecule $p$ respectively.
    Applying the hopping operator we define a state
    ${\ket{\zeta^{(c)}(0)} = \hat V_{pq}^{c} \ket{\Psi_{\aset^\prime \cup \{q\}}^0\Phi_p^0}}$.
    After hopping, the active sector becomes the 
    set of molecules $\aset^\prime \cup \{p\}$.
    We may then time evolve this state:
    \begin{displaymath}
    \ket{\zeta^{(c)}(t)}=e^{-i\left(H^{HTC}_{\aset^\prime \cup \{p\}}+H^{H}_{q}\right)t/\hbar}\ket{\zeta^{(c)}(0)},
    \end{displaymath}
    by numerical integration of the Schrodinger equation below.
    The time-domain response function is then: $\hopresp^{(c)}(t)=\braket{\zeta^{(c)}(t)}{\zeta^{(c)}(0)}$.
    
    The operator $\hat V_{pq}^{\sigma} $  swaps the electronic states of molecules $p$ and $q$, and leaves their vibrational states unchanged.  As a result,
    the vibrational state of molecule $q$, which becomes charged (thus optically inactive)  after the hopping, remains entangled with the state of all of the active molecules (except molecule $p$).
    Because of this, we cannot factorize $\ket{\zeta^{(c)}(0)}$ into active and charged sectors,  and so we have to perform the time evolution in the combined space of all molecules, $\aset^\prime \cup \{p, q\}$.
    
    In the following, we provide some technical details of our numerical implementation of the above approach, which uses the permutation symmetry of all molecules  not involved in the hopping process.

    \paragraph*{Hamiltonian.}
    We focus on  $H^{HTC}_{\aset^\prime \cup \{p\}}$, as the calculation of $H^{H}_{q}$ (acting on a single molecule) is trivial.
    First, consider the relevant Hilbert space for $H^{HTC}_{\aset^\prime \cup \{p\}}$, which we denote $\mathcal{H}^{HTC}_{\aset^\prime \cup \{p\}}$. 
    We define $\mathcal{H}^{HTC}_{\aset^\prime,N_{ex}}$ as the the permutation symmetric subspace with $N_{ex}$ excitations distributed between the cavity mode and the subset of active molecules $\aset^\prime$ (which excludes molecules $p,q$).  We can then write the Hilbert space as
    \begin{multline*}
    \mathcal{H}^{HTC}_{\aset^\prime \cup \{p\}}=\mathcal{H}^{HTC}_{\aset^\prime,N_{ex}}\otimes \{\ket{\da_p,\nu_p}\} \\\oplus {\mathcal{H}^{HTC}_{\aset^\prime,N_{ex}-1}\otimes \{\ket{\ua_p,\nu_p}\}},
    \end{multline*}
    where $\nu_p$ is the number of vibrational excitations on molecule $p$.
    Given this structure, it is helpful to divide the Hamiltonian $H^{HTC}_{\aset^\prime \cup \{p\}}$ into blocks in the two subspaces  
    using projection operators $\hat P_{\sigma_p}=\ket{\sigma_p}\bra{\sigma_p}$:
    \begin{multline}
    \label{eq:HTCsplitspace}
    H^{HTC}_{\aset^\prime \cup \{p\}}
    =
    \hat P_{\da_p} \left(
      H^{HTC}_{\aset^\prime, N_{ex}} + \wv\bd_p\bn_p
    \right)
   \\
   +
   \hat P_{\ua_p} \left(
     H^{HTC}_{\aset^\prime, N_{ex}-1}
     + \wo + \wv\left[\bd_p\bn_p + \lam(\bd_p + \bn_p)\right]
   \right)
   \\
   +
   \frac{\wrr}{\sqrt{N}}(\an\sd_p + \ad\sn_p).
    \end{multline}
    Here, $H^{HTC}_{\aset^\prime, N_{ex}}$ is the HTC Hamiltonian with $N_{ex}$ excitations among the cavity mode and the active molecules $\aset^\prime$.  This can be written using the method described in Sec.~\ref{sec:htc-hamilt-perm}.  The last line of Eq.~\eqref{eq:HTCsplitspace} has the effect of connecting the two subspaces.
    
    \paragraph*{Initial state.}
    Having defined the Hamiltonian, we need next to specify how to find the initial state $\ket{\zeta^{(c)}(0)}= \hat V_{pq}^{c} \ket{\Psi_{\aset^\prime \cup \{q\}}^0\Phi_p^0}$.  The original state $\ket{\Psi_{\aset^\prime \cup \{q\}}^0}$ can be found by using the Lanczos algorithm, while $\ket{\Phi_p^0}$ can be written directly.

    As described above, the pre-hopping state lives in the space $\mathcal{H}^{HTC}_{\aset^\prime \cup \{q\}}\otimes\mathcal{H}^{H}_{p}$,
    while the state after the hopping $\ket{\zeta^{(c)}(0)}$ lives in the space $\mathcal{H}^{HTC}_{\aset^\prime \cup \{p\}}\otimes\mathcal{H}^{H}_{q}$
    (where $\mathcal{H}_H^{q}$ denotes the Hilbert space of a single charged molecule).
    Since all molecules are identical, 
    instead of interchanging the electronic states of the hopping molecules we can equivalently swap the labeling of the molecules and their vibrational states.
    As such, to obtain the vector $\ket{\zeta^{(c)}(0)}$, we can interchange the vibrational states of the charged molecule with that of the active molecule in the appropriate electronic state manifold:
    $\ket{\sigma_p,\nu_p}\ket{D_q,\mu_q} \to \ket{\sigma_p,\mu_q}\ket{D_q,\nu_p}$
    for all $\nu,\mu$. This can be performed using the indexing functions described above to determine the effect of adding a molecule with a given vibrational state to the active set $\aset^\prime.$

    \paragraph*{Numerical integration.}
    \label{sec:integration}
    We use the Runge-Kutta algorithm to integrate the Schrodinger equation,
    $i\hbar\frac{d}{dt}\ket{\zeta^{(c)}(t)}={(H - i\kappa/2)}\ket{\zeta^{(c)}(t)}$,
    with the given initial condition, $\ket{\zeta^{(c)}(0)}$, calculated as described above.
    Here, $H=H^{HTC}_{\aset^\prime \cup \{q\}} + H^{H}_{p}$ and $\kappa$ is a small broadening added so that the state 
    and hence correlation $\hopresp^{(c)}(t)$
    decays with time
    producing a smooth Fourier transform $\hopresp^{(c)}(\omega)$.
    
    \subsubsection{Relation to vibrational density matrix elements }
    
    It is instructive to see how the response function, $\hopresp^{(c)}(t)$, can also be directly related to the reduced density matrices mentioned above.  We discuss this here.
    
    Noting that  the hopping operators can be written using a resolution of identity in the vibronic basis states,
    ${\hat V_{pq}^c=\sum_{\mu\nu}\ket{\sigma(c)_{p} \mu_p,D_q \nu_q}\bra{ D_p \mu_p,\sigma(c)_q \nu_q}}$, where $\sigma(L,H)=\da,\ua$ respectively.
    Using both the Holstein--Tavis--Cummings and Holstein Hamiltonians, the response function 
    $\hopresp^{(c)}(t)$
    can be written as, 
    \begin{multline}
    \hopresp^{(c)}(t)=
    \sum_{\mu\nu\mu^\prime\nu^\prime}
    \braket{\Phi_p^0,\Psi_{\aset^\prime \cup \{q\}}^0}{D_p \nu^\prime_p,\sigma(c)_q \mu^\prime_q} \times
    \\
    \mel{\sigma(c)_{p} \nu^\prime_p,D_q \mu^\prime_q}{%
    e^{-i\left(H^{HTC}_{\aset^\prime \cup \{p\}} + H^{H}_{q}\right)t/\hbar}
     }{\sigma(c)_{p} \mu_p,D_q \nu_q}
    \\
    \times
    \braket{ D_p \mu_p,\sigma(c)_q \nu_q}{\Phi_p^0,\Psi_{\aset^\prime \cup \{q\}}^0}.
    \end{multline}
    We can factorize this expression as follows:
    \begin{align}
    \hopresp^{(c)}(t)
    &=\sum_{\mu\nu\mu^\prime\nu^\prime}
    \chi^{\sigma(c)}_{\nu\mu\mu^\prime\nu^\prime}(t) \times \chi^D_{\mu\nu\nu^\prime\mu^\prime}(t),\\
    \chi^{\sigma}_{\mu\nu\mu^\prime\nu^\prime}(t)
    &\equiv
    \braket{\Psi^0}{\sigma \nu^\prime}
    \mel{\sigma \mu^\prime}{e^{-iH^{HTC}t/\hbar}}{\sigma \nu}
    \braket{\sigma \mu}{\Psi^0},
    \nonumber\\
    \chi_{\mu\nu\mu^\prime\nu^\prime}^{D}(t)
    &\equiv
    \braket{\Phi^0}{D \nu^\prime}
    \mel{D \mu^\prime}{e^{-iH^{H}t/\hbar}}{D \nu}
    \braket{D \mu}{\Phi^0}.
    \nonumber
    \end{align}
    We have suppressed the molecule labels, since only one of $p,q$ appears in each factor.
    The behavior on the doubly occupied molecule, 
    $\chi^{D}_{\mu\nu\mu^\prime\nu^\prime}(t)$ is straightforward to obtain, as this single molecule evolves on its own.  We thus have
    \begin{displaymath}
    \chi^{D}_{\mu\nu\mu^\prime\nu^\prime}(t)
    =\sum_{s}
    \mathcal{D}^{\dagger}_{0,\nu^\prime}
    \mathcal{D}_{\mu^\prime,s}
    e^{-i s\wv t} 
    \mathcal{D}^{\dagger}_{s,\nu}
    \mathcal{D}_{\mu,0},
    \end{displaymath}
    with
    ${\mathcal{D}_{\mu,s}=\braket{\mu}{\Phi^s}}$,
    and $s$ counts the number of vibrational excitations.
    This can also be written as,
    ${
     \chi^{D}_{\mu\nu\mu^\prime\nu^\prime}(t)=\sum_{s} e^{-i s \wv t} (\rho_D^s)_{\mu\nu} (\rho_D^s)^{\dagger}_{\mu^\prime\nu^\prime}
    }$,
    where 
    $
    (\rho^{s}_{D})_{\mu\mu^\prime}=
      \braket{D \mu }{\Phi^0}
      \braket{\Phi^s}{D \mu^\prime}
    $
    is the doubly-occupied sector equivalent of ${\rho_\sigma^r}$ defined in Sec.~\ref{sec:cond-reduc-vibr}
    
    \begin{widetext} 
    The behavior of the optically active molecule,
    $\chi_{\mu\nu\mu^\prime\nu^\prime}^{\sigma}(t)$
    can then be obtained from the expressions $\rho_{\sigma}^r$ given in Sec.~\ref{sec:cond-reduc-vibr}.
    We may write a resolution of identity 
    $\mathbb{1}=\sum_{r}\ket{\Psi^r}\bra{\Psi^r}$ in terms of the eigenstates $\ket{\Psi^r}$ of the HTC model, with energies $E^{HTC}_r$.
    Inserting this
    after the exponential term in the expression for $\chi_{\mu\nu\mu^\prime\nu^\prime}^{\sigma}(t)$,
    we obtain
    \begin{align*}
    \chi^{\sigma}_{\mu\nu\mu^\prime\nu^\prime}(t)
    &=\sum_{r}
    \braket{\Psi^0}{\sigma \nu^\prime}
    \braket{\sigma \mu^\prime}{ \Psi^r}
    e^{-i E^{HTC}_r t/\hbar}
    \braket{\Psi^r}{\sigma \nu}
    \braket{\sigma \mu}{\Psi^0}
    =
    \sum_{r} e^{-i E^{HTC}_r t/\hbar}
    (\rho_{\sigma}^r)_{\mu\nu}
    (\rho_{\sigma}^r)^{\dagger}_{\mu^\prime\nu^\prime}.
    \end{align*}
    The remaining sums and convolutions can in principle be evaluated numerically.
    \end{widetext}
    
    \subsection{Evaluating spectral weights}
    \label{sec:eval-spectr-weights}
    
    In Figure~\ref{fig:abs}, we plotted the system-size dependence of the probability
    ${\matelem^{LP_1(H)}}$.  For calculating this, we used the expression
    \begin{equation}
     \matelem^{l_{j,k}(c)}=
     \left|
     \text{Tr}\left(
        \rho^{j}_{\sigma}[\rho_D^{k}]^T
     \right)
     \right|^2, 
    \end{equation}
    in terms of the quantities
    $(\rho^{j}_{\sigma})_{\nu\mu}=\braket{\Psi^j}{\sigma \mu}\braket{\sigma \nu}{\Psi^0}$, 
    and
    $(\rho_D^{k})_{\mu\nu}=\braket{\Phi^k}{D \nu}\braket{D \mu}{ \Phi^0}$
    introduced above.

    \section{Evolution of Gaussian fitting parameters with excitation density}
    \label{sec:fitt-param-evol}

    \begin{figure}[hptb]
      \centering
    \includegraphics[width=1\columnwidth]{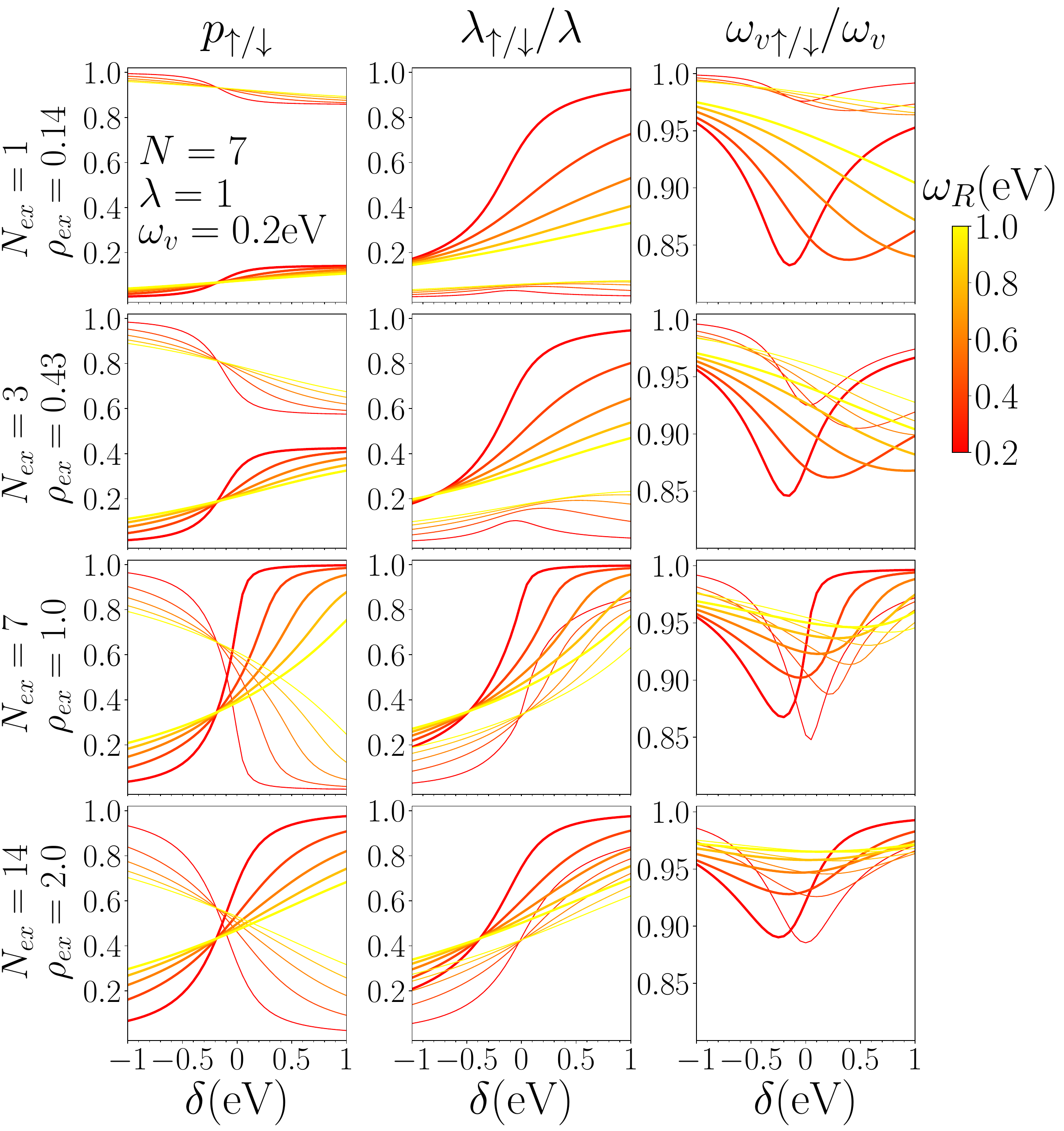}
    \caption{ Fitting parameters for the reduced 
      density matrix $\rho^0_{\sigma}$ of the ground state with $N_{ex}$ excitations. Plotted vs $\delta$ at various $\wrr$ as shown by the colorscale.  Each row shows a different values of $N_{ex}$.
      Left: Electronic state probability $p_{\sigma}$. Middle:
      conditional displacement $\lam_{\sigma}$. Right: conditional frequency
      $\wv_{\sigma}$. Unexcited state ($\sigma=\da$) parameters are thin lines, and
      excited state ($\ua$) are thicker lines.  Other parameters $N=7,\lam=1,\wv=0.2$eV.}
    \label{fig:effpar}
    \end{figure}
    
    The results in Sec.~\ref{sec:param-evol-with} discussed the evolution of $\lam_\sigma$ for a special case where $\rho_{ex}=1$.  Here we discuss how the behavior evolves with changing $\rho_{ex}$. Figure~\ref{fig:effpar} shows all three fitting parameters,
    $p_{\sigma}$, ${\wv}_{\sigma}$, $\lam_{\sigma}$, with $\delta$ and $\wrr$ dependence as in Fig.~\ref{fig:fit}(b), but with each row corresponding to a different excitation density, $\rho_{ex}$.

    We first we discuss the left-hand column, $p_{\sigma}$. By definition, $p_{\ua} + p_{\da} =1$, so we focus on the evolution of $p_{\ua}$.  At large negative detuning, the polariton state becomes purely photonic, so $p_{\ua} \to 0$.
    The behavior at large positive detuning depends on $\rho_{ex}$.
    For $\rho_{ex}<1$, there are insufficient excitations for all molecules to be excited, so  $p_{\ua}\simeq\rho_{ex}$. For
    $\rho_{ex} > 1$, there will always be a non-zero photon field,  causing hybridization between excitonic states so $p_{\sigma}<1$. When $\rho_{ex} \gg 1$, as discussed in  Sec.~\ref{sec:novib-large-rho}, this leads to $p_{\sigma} \to 1/2$.

    Regarding $\lam_\sigma$, the behavior at $\rho_{ex}=1$ was discussed in Sec.~\ref{sec:param-evol-with} and Sec.~\ref{sec:pert-theory-at}.
    At large $\wrr$, the behavior remains similar for other values of $\rho_{ex}$. 
    For small $\wrr$, the behavior also remains similar when $\rho_{ex}>1$ or $\td$ is negative.  
    For $\td>0$, i.e. $\delta>-\lam^2\wv=-0.2$ and $\rho_{ex}<1$, the behavior does change as the ground state here is no longer a fully excited state.  As such (as discussed at the end of Sec.~\ref{sec:pert-theory-at}) one finds $\lam_{\da} \to 0, \lam_{\ua}\to \lam$.
    
    Regarding $\wv_\sigma$, there is relatively little dependence on $\delta$ or $\rho_{ex}$ (note the scale in the right column of Fig.~\ref{fig:effpar}).  The slight reduction below one means the probability distributions are slightly broadened.
    
    
%

    \end{document}